\documentclass{pasj00}

\begin{document}
\SetRunningHead{Y. Takeda et al.}{C and O abundances of Pleiades F--G stars}
\Received{2016/09/06}
\Accepted{2016/10/13}

\title{Photospheric carbon and oxygen abundances \\ 
of F--G type stars in the Pleiades cluster
\thanks{Based on data collected by using the 1.5-m Telescope
of Gunma Astronomical Observatory, and those with the 1.88-m Telescope
of Okayama Astrophysical Observatory (obtained from SMOKA, operated
by the Astronomy Data Center, National Astronomical Observatory of Japan).}
}

%

\author{
Yoichi \textsc{Takeda}\altaffilmark{1,2},
Osamu \textsc{Hashimoto}\altaffilmark{3},
 and
Satoshi \textsc{Honda}\altaffilmark{4}
}
\altaffiltext{1}{National Astronomical Observatory, 2-21-1 Osawa, 
Mitaka, Tokyo 181-8588}
\email{takeda.yoichi@nao.ac.jp}
\altaffiltext{2}{SOKENDAI, The Graduate University for Advanced Studies, 
2-21-1 Osawa, Mitaka, Tokyo 181-8588}
\altaffiltext{3}{Gunma Astronomical Observatory, 6860-86 Nakayama, 
Takayama, Agatsuma, Gunma 377-0702}
\altaffiltext{4}{Nishi-Harima Astronomical Observatory, Center for Astronomy,\\
University of Hyogo, 407-2 Nishigaichi, Sayo-cho, Sayo, Hyogo 679-5313}

%


\maketitle

\begin{abstract}
In order to investigate the carbon-to-oxygen ratio of the young open
cluster M~45 (Pleiades), the C and O abundances of selected 
32 F--G type dwarfs (in the effective temperature range of 
$T_{\rm eff} \sim$~5800--7600~K and projected rotational velocity range 
of $v_{\rm e}\sin i \sim$~10--110~km~s$^{-1}$) 
belonging to this cluster were determined by applying the synthetic 
spectrum-fitting technique to C~{\sc i} 5380 and O~{\sc i} 6156--8 lines.
The non-LTE corrections for these C~{\sc i} and O~{\sc i} lines   
were found to be practically negligible (less than a few hundredths dex).
The resulting C and O abundances (along with the Fe abundance) turned out nearly 
uniform without any systematic dependence upon $T_{\rm eff}$ or $v_{\rm e}\sin i$.
We found, however, in spite of almost solar Fe abundance ([Fe/H]~$\sim 0$), 
carbon turned out to be slightly subsolar ([C/H]~$\sim -0.1$) while oxygen 
slightly supersolar ([O/H]~$\sim +0.1$). This lead to a conclusion that 
[C/O] ratio was moderately subsolar ($\sim -0.2$) in the primordial gas 
from which these Pleiades stars were formed $\sim$~120--130~Myr ago. 
Interestingly, similarly young B-type stars are reported to show just the same result 
([C/O]~$\sim -0.2$), while rather aged ($\sim$~1--10~Gyr) field F--G stars of 
near-solar metallicity yield almost the solar value ([C/O]~$\sim 0$) on the average.
Such a difference in the C/O ratio between two star groups of distinctly 
different ages may be explained as a consequence of orbit migration 
mechanism which Galactic stars may undergo over a long time.
\end{abstract}

%


\section{Introduction}

The abundances of carbon and oxygen in the Galaxy are currently attracting hot
interest of astrophysicists, because they play significant roles in the chemical 
characteristics of celestial objects such as stars and planets. 
For example, carbon-dominated rocky planets (rather than oxygen-dominated ones
such as those existing in the solar system) are expected to form if C/O ratio 
of the circumstellar gas is sufficiently high to satisfy $N$(C)/$N$(O)$>0.8$ 
(e.g., Bond et al. 2010), which corresponds to [C/O] $>$ +0.16 if Asplund 
et al.'s (2009) solar abundances of $A_{\odot}$(C) = 8.43 and $A_{\odot}$(O) 
= 8.69 are adopted.\footnote{In this paper, we define the logarithmic number 
abundance for element X as $A$(X)$ \equiv \log [N({\rm X})/N({\rm H})] + 12$, 
where $N$(X) and $N$(H) is the number density of element X and H (hydrogen), respectively. 
Similarly, we use the usual notation for the differential logarithmic abundance 
ratio relative to the Sun: [X/Y]$_{*} \equiv$ [X/H] $-$ [Y/H] $\equiv$  
$[A_{*}({\rm X}) -  A_{\odot}({\rm X})] -  [A_{*}({\rm Y}) -  A_{\odot}({\rm Y})]$,
where the subscripts ``*'' and ``$\odot$'' mean ``object'' and ``Sun'', respectively.
}
Accordingly, the primary issue is whether such stars with appreciably supersolar 
[C/O] ratio ([C/O] $\gtsim 0.2$) exist or not, which can be checked by examining
spectroscopically determined C and O abundances of solar-type stars.  
Generally speaking, rather metal-rich stars would be promising for such candidates, 
because [C/O] values observed in F--G stars in the Galactic disk tend to rise over 
zero at [Fe/H]~$\gtsim 0$ as a result of slower decreasing rate of [C/Fe] than [O/Fe] 
with an increase of [Fe/H] (see, e.g., Fig.~7 of Gustafsson et al. 1999, or 
Fig.~6d in Takeda \& Honda 2005). 

However, despite that several groups challenged this problem recently, 
considerably discordant results have been reported: Delgado Mena et al. (2010) as 
well as Petigura and Marcy (2011) argued that a significant fraction ($\sim$~20--30\%) 
of nearby FGK stars have $N$(C)/$N$(O)$>0.8$ (even $N$(C)/$N$(O)$>1$ is found for 
$\gtsim$~5\% of the sample). On the other hand, Nissen et al. (2014) criticized 
the consequences of these two studies as being questionable (presumably due to the 
difficult nature of adopted forbidden [O~{\sc i}] 6300 line which suffers a blending 
with Ni~{\sc i} line and has a different parameter-sensitivity in comparison to 
the high-excitation C~{\sc i} lines also used by them), and concluded based on high-excitation 
C~{\sc i} 5052/5380 and O~{\sc i} 7771--5 lines that very few F--G stars in the 
solar neighborhood exceed the critical $N$(C)/$N$(O) limit of 0.8 even at the highest 
metallicity of [Fe/H] = +0.4. Furthermore, Nakajima and Sorahana (2016) very recently 
examined the distribution of C-to-O abundance ratio of M dwarfs, and lent support
rather to the latter conclusion of Nissen et al. (2014). In any event, more 
observational studies would be needed to settle this controversy.

In this connection, investigating the C and O abundances of stars belonging 
to an open cluster is of significance, since (1) they represent the original 
composition of the gas at a well-established time point in the Galactic history
and (2) fairly accurate abundances are determinable by comparing/averaging
the results of many members by making use of the chemical homogeneity.
Takeda et al. (2013; hereinafter referred to as Paper~I) once studied
the C and O abundances of F--G stars in the Hyades cluster (age of $\sim$~625~Myr;
cf. Perryman et al. 1998). 
Interestingly, as remarked in Sect.~6.2 therein, a slightly subsolar 
[C/O]~($\sim -0.1$) was obtained, despite that this cluster is somewhat metal-rich 
([Fe/H]~$\sim$0.1--0.2) and expected to have a trend of [C/O]~$\gtsim 0$ according 
to the [C/O]--[Fe/H] relation mentioned above (derived from field FGK stars with 
ages of $\sim$1--10~Gyr). Does any different tendency holds for the [C/O] 
ratios of rather young cluster stars? 

In view of this consideration, it is meaningful to check the C and O abundances of 
another well-known open cluster M~45 (Pleiades), which has an age of $\sim$~120--130~Myr 
(cf., Basri, Marcy, \& Graham 1996; Bonatto, Bica, \& Girardi 2004) and 
thus further younger than Hyades.
Unfortunately, available previous investigations of C and O abundances in the Pleiades 
stars are not so many, despite that various spectroscopic studies of other elements 
have been published (see, e.g., Table~I of Gebran \& Monier 2008 or Sect.~1 of 
Funayama et al. 2009) where the metallicity ([Fe/H]) was reported to be nearly solar.\\
--- Friel and Boesgaard (1990) determined the C abundances for 12 Pleiades 
dwarfs ($T_{\rm eff} \sim$~5900--6900~K) by using C~{\sc i} 7771--9 and 
C~{\sc i} 6587 lines and derived an almost near-solar (or very slightly subsolar) 
tendency for [C/H] ($-0.06$).\\
--- Boesgaard (2005) reported that the C and O abundances of 20
late-F and early-G stars ($T_{\rm eff} \sim$~5600--6200~K) determined from
C~{\sc i} 7111--9 and O~{\sc i} 7771--5 lines were almost solar 
(the mean abundances were $\langle$[C/H]$\rangle$ = $-0.02$ and
$\langle$[O/H]$\rangle$ = $+0.01$).\\ 
--- Gebran and Monier (2008) carried out abundance studies for 21 stars (16 
A-type and 5 F-type) of the Pleiades cluster for 18 elements including C and O.
By using a spectrum-fitting technique applied to a number of C~{\sc i} and O~{\sc i} 
lines, they found that C and O tend to be underabundant relative to the Sun,
where the extent of deficiency is larger for A stars ($-0.5 \ltsim$~[C/H]~$\ltsim$~+0.2 
and $-0.4 \ltsim$~[O/H]~$\ltsim$~+0.2 ) than F stars ($-0.1 \ltsim$~[C/H]~$\ltsim$~0.0 
and $-0.3 \ltsim$~[O/H]~$\ltsim$~0.0 ). We should bear in mind, however,
that light-element abundances of A-type stars may suffer chemical anomalies
due to the diffusion process in the outer layer.

These previous studies suggest that C and O abundances of Pleiades are not 
so much different from the solar composition and thus precise determination 
of the C/O ratio in this young cluster needs a careful analysis.  
Motivated by this situation, we decided to contend with this problem by conducting 
new determinations of C and O abundances for selected 32 Pleiades F--G type stars, 
with an aim of clarifying the characteristics regarding [C/H], [O/H], and [C/O] of 
this young cluster. This is the purpose of this investigation.

The remainder of this article is organized as follows. 
After describing our observational data in section 2, 
we explain the assignments of atmospheric parameters in section 3. 
The procedure of our abundance determination based on the spectrum-fitting 
method is illustrated in section 4. Section 5 discusses the
consequence of our analysis, where the resulting abundances of C and O 
along with the C/O ratios of the Pleiades stars are examined in various respects. 
In addition, two supplementary sections are prepared to justify the adopted  
reference solar oxygen abundance derived from the O~{\sc i} 6156--8 lines 
(appendix~1) and to briefly report the Li abundances of 7 Pleiades stars
among our sample (appendix~2).

\section{Observational data} 

A total of 32 F--G dwarfs of the Pleiades cluster were selected as the 
sample stars of this study. Regarding 25 stars among them, we 
made use of the high-dispersion spectral data (with a spectral resolving power
of $R\sim 40000$) in the green--red region originally obtained in 2006--2007 
by Funayama et al. (2009) and employed for their metallicity study of
Pleiades stars, which are archived as raw (unprocessed) data at 
Gunma Astronomical Observatory (GAO) and Okayama Astrophysical Observatory (OAO). 
See Sect. 2 in Funayama et al. (2009) regarding the observational details
of these data.

In addition, we newly carried out spectroscopic observations of 7 stars 
in the 2015--2016 season by using GAOES (Gunma Astronomical Observatory 
Echelle Spectrograph) installed at the Nasmyth Focus of the 1.5~m reflector 
of GAO, by which $R \sim 40000$ spectra (corresponding to the slit width of
2$''$) covering the wavelength range of 5340--6580~$\rm\AA$ were obtained.

The data reduction for all these 32 stars (bias subtraction, flat-fielding, 
aperture-determination, scattered-light subtraction, 
spectrum extraction, wavelength calibration, and continuum-normalization) 
was performed using the ``echelle'' package of IRAF.\footnote{IRAF is 
distributed by the National Optical Astronomy Observatories, 
which is operated by the Association of Universities for  Research  
in Astronomy, Inc., under cooperative agreement with the National 
Science Foundation.} Generally, a number of spectral frames (sometimes
those obtained over several different nights)\footnote{Regarding MSK~65 (= HII~761),
since the radial velocity of this star was found to be considerably variable 
with a short time scale, we had to correct the wavelength scale of each spectrum 
to the laboratory frame before co-addition.}
for each star were co-added in order to improve the data quality.
The typical S/N ratio of the resulting spectra is
around $\sim 100$ (i.e., $\sim$~40--50 for the worst cases while 
$\sim$~150 for the best cases) . The basic information of the finally adopted 
observational data for our 32 sample stars (along with their photometric data
taken from the SIMBAD database) is summarized in table~1.

Besides, since we selected Procyon (F5~III--IV) and Sun (Moon) as the reference stars,
their spectra were taken from Takeda et al.'s (2005a) spectrum database
of 160 FGK stars which were obtained at OAO.\footnote{Since there was a serious defect 
in Takeda et al.'s (2005a) Moon spectrum around 5375--5377~$\rm\AA$, we had to discard
this spectrum portion in the fitting process, by which Mn abundance could not
be determined for the Sun (cf. figure~3).}

\section{Atmospheric parameters}

The fundamental atmospheric parameters [the effective temperature 
($T_{\rm eff}$), logarithmic surface gravity ($\log g$), microturbulence ($\xi$)] 
of each 32 Pleiades star were determined by making use the fact that all 
the objects (member stars of the Pleiades cluster) are known to have almost the 
same age ($\sim$~120--130~Myr) and the solar abundance (cf. section~1). 
We adopted the solar-metallicity isochrones\footnote{
Available at $\langle$http://stev.oapd.inaf.it/cgi-bin/cmd$\rangle$.}
computed by Bressan et al.'s (2012, 2013) PARSEC code (version 1.2S) 
corresponding to the age of 125~Myr,\footnote{Actually, the parameter-determination 
procedure described in this section was performed for two different isochrones
with ages of 100~Myr and 150~Myr. Since they were found to be hardly discernible 
from each other in the considered $(B-V)_{0}$ range, we simply averaged the both 
results to derive the solution corresponding to 125~Myr.} 
in which the values of $M_{V0}$ (absolute visual magnitude), $(B-V)_{0}$ (unreddened 
color), $T_{\rm eff}$, and $g$ are tabulated in terms of the stellar mass.
These data suffice us to express $M_{V0}$, $T_{\rm eff}$, and $\log g$
in terms of $(B-V)_{0}$, which was derived from the observed
$B-V$ color (taken from the SIMBAD database) along with the typical color excess 
of $E_{B-V} = 0.03$~mag (or the extinction is $A_{V} = 3.1E_{B-V} = 0.093$~mag) 
for the Pleiades cluster (see Table~3 of Breger 1986).
The $V_{0} (\equiv V - A_{V})$ vs. $(B-V)_{0}$ plots for the program stars are compared with
the isochrones (where $M_{V0}$ was reduced to $V_{0}$ by using the distance of 
136.2~pc; cf. Melis et al. 2014) in figure 1a.
The final $T_{\rm eff}$--$(B-V)_{0}$ and $\log g$--$(B-V)_{0}$ relations
we adopted for our Pleiades stars are shown (by solid lines) in figures 1b and 1c, 
respectively, where other empirical formula occasionally used for FGK dwarfs are also shown
(by dashed lines) for comparison.  

Regarding the microturbulence, we employed the analytical formula for $\xi$
in terms of $T_{\rm eff}$ and $\log g$ derived in Paper~I [cf. 
equation~(1) therein], by which $\xi$ can be expressed as a function of $(B-V)_{0}$
by using the relations specific for Pleiades stars mentioned above. 
Such derived $\xi$ vs. $(B-V)_{0}$ relation we adopted is illustrated in figure 1d
(solid line), where similar relations based on alternative empirical 
$\xi(T_{\rm eff}, \log g)$ formulas of Nissen (1981) and Edvardsson et al. (1993) 
are also shown for comparison (dashed lines). 

The finally adopted values of $T_{\rm eff}$, $\log g$, and $\xi$ for the
program stars are given in table 2 (columns 3--5). 
These isochrone-based parameters are reasonably consistent with those derived 
spectroscopically by Funayama et al. (2009) using Fe~{\sc i} and Fe~{\sc ii} lines,
as can be recognized in figures 2a--2c.
As to the atmospheric parameters for Procyon and the Sun, we employed the same values
as used in Takeda et al. (2005b; cf. table~1 therein).

The adopted $(B-V)_{0}$ values, on which our parameters are based, may have inevitable 
ambiguities of several hundredths mag (as estimated from the dispersion of $E_{B-V}$ 
in Pleiades stars shown in Fig.~4 of Breger 1986), which are translated to
uncertainties in $T_{\rm eff}$ by $\sim \pm 200$~K (cf. figure~1b). Regarding errors 
in $\log g$ and $\xi$, we assume $\sim \pm 0.1$~dex and $\sim \pm 0.5$~km~s$^{-1}$
as done in Paper~I, respectively, which may be reasonable estimates as judged from 
figures~1c and 1d. 

\section{Abundance determination}

\subsection{Spectrum-fitting analysis}

The model atmosphere for each star was constructed by two-dimensionally 
interpolating Kurucz's (1993) ATLAS9 model grid (models with convective overshooting) 
with respect to $T_{\rm eff}$ and $\log g$ determined in section~3, 
where we exclusively applied the solar-metallicity models. 

Our abundance determination procedure is basically the same as adopted 
in Paper~I (see section~5 therein), which consists of three consecutive steps: 
(i) synthetic spectrum fitting to derive provisional abundances for the important 
elements in the relevant spectrum range, (ii) inverse calculation of the equivalent 
width for the line (or the blended feature) of C or O in question by using the 
fitting-based abundance solutions, and (iii) analysis of such established equivalent 
widths to derive the final abundances or abundance errors due to uncertainties 
in atmospheric parameters.

For the purpose of determining C and O abundances, we decided to employ 
C~{\sc i} 5380.325 line (instead of C~{\sc i} 7111-9 lines adopted in Paper~I,
which are unavailable in our data) and O~{\sc i} 6156--8 lines (which were also used in Paper~I),
since these two line features are available in all the spectra at our hand (cf. table~1).
Accordingly, we selected two wavelength regions: (a) 5375--5390~$\rm\AA$ region 
(comprising lines of C, Ti, Mn, and Fe), and
(b) 6150--6167~$\rm\AA$ region (comprising lines of O, Na, Si, Ca, Fe, and Ni). 
The atomic data of the spectral lines were taken from VALD database (Ryabchikova et al. 
2015) for former 5375--5390~$\rm\AA$ region, while from  Kurucz and Bell's (1995) 
compilation for the latter 6150--6167~$\rm\AA$ region (in order to maintain
consistency with Paper~I where the same O~{\sc i} 6156--8 lines were used).
Although all the atomic lines contained in each database were included
for each wavelength region, only the selected data for especially important lines 
are summarized in table~3.

As to the macroscopic broadening function (to be convolved with the intrinsic
spectrum to simulate the synthetic stellar spectrum), we adopted the rotational 
broadening function (see, e.g., Gray 2005) with the limb-darkening coefficient of 
$\epsilon = 0.5$, which means that $v_{\rm e}\sin i$ (projected rotational velocity) 
is the control parameter for adjusting the line width in the fitting.
Note here that, since neither instrumental broadening nor macroturbulence
is explicitly included in our modeling, the resulting $v_{\rm e}\sin i$
solution should not always be regarded as equivalent to true (projected)
rotational velocity: That is, in case of sharp lines (e.g., $\ltsim 10$~km~s$^{-1}$), 
our $v_{\rm e}\sin i$ solution may lead to some overestimation (though such 
a case is rather exceptional because rotational broadening is dominant for most 
of our sample stars).
Accordingly, the parameters to be varied to accomplish the best fit 
between the theoretical and observed spectrum by using Takeda's (1995) numerical 
technique are (i) the abundances of relevant elements, (ii) $v_{\rm e}\sin i$,
and (iii) $\Delta\lambda$ (radial velocity shift).

The convergence of the fitting solutions turned out fairly successful  
for most of the cases. How the theoretical spectrum for the converged 
solutions fits well with the observed spectrum for each of the 32 Pleiades stars 
(as well as Procyon and Sun/Moon) is displayed in figure~3 (5375--5390~$\rm\AA$ 
fitting, where C, Ti, Mn, and Fe abundances were varied) and figure~4 
(6150--6167~$\rm\AA$ fitting, where O, Na, Si, Ca, Fe, and Ni abundances were varied).
The finally resulting values of elemental abundances ($A$) as well as $v_{\rm e}\sin i$
derived from the fitting procedure for each region are presented in tableE.txt 
(on-line material).

\subsection{Equivalent widths and abundance uncertainties}

Next, as done in Paper~I, we computed the equivalent widths for 
the C~{\sc i} 5380.325 line ($W_{5380}$)\footnote{Actually, several other weaker 
C~{\sc i} lines exist in the neighbourhood of this C~{\sc i} 5380.325 line (cf. table~3),
among which the C~{\sc i} 5380.227 line (of higher excitation) may not be negligible.
However, since $W$(5380.227) turned out only $\sim$~20\% compared with $W$(5380.325),
as demonstrated in figure~5a, it can not make any important contribution.
So, we considered only the C~{\sc i} 5380.325 line in evaluating the non-LTE correction
($\Delta_{5380}$).} and the O~{\sc i} 6158 feature ($W_{6158}$; 
comprising three components) based on these fitting-based abundances.
Then, the non-LTE abundances [$A^{\rm N}$(C) and $A^{\rm N}$(O)] and the relevant non-LTE corrections 
($\Delta_{5380}$ and $\Delta_{6158}$; where $\Delta \equiv A^{\rm N} - A^{\rm L}$) 
were derived based on these $W_{5380}$ and $W_{6158}$, where the non-LTE effect for 
the C and O lines was taken into account as done by Takeda and Honda (2005).
The resulting equivalent widths and  non-LTE abundances as well as corrections
for the sample stars are also given in tableE.txt (on-line material). 

The uncertainties in $A$(C) and $A$(O) due to errors in the 
adopted atmospheric parameters were estimated by repeating the analysis 
on the $W$ values while perturbing the standard parameters interchangeably 
by $\pm 200$~K in $T_{\rm eff}$, $\pm 0.1$~dex in $\log g$, and $\pm 0.5$~km~s$^{-1}$ 
in $\xi$ (which are considered to be typical magnitudes of ambiguities; 
cf. the last paragraph in section~3).

We also evaluated errors due to random noises of the observed spectra by 
estimating S/N-related uncertainties in the equivalent width ($\delta W$) 
by invoking the relation derived by Cayrel (1988),
$\delta W \simeq 1.6 (w \delta x)^{1/2} \epsilon$,
where $\delta x$ is the pixel size ($\simeq 0.025$~$\rm\AA$), $w$ is the
width of a line (which we assumed to be $\lambda v_{\rm e}\sin i/c$, 
where $c$ is the velocity of light), and $\epsilon \equiv ({\rm S/N})^{-1}$.
This approach is known to yield abundance errors of almost the same order 
of magnitude as those derived by extensive numerical simulations based on 
a large number of mock spectra with artificial noises (cf. Takeda \& Honda 2015). 
We thus determined the abundances for each of the perturbed 
$W_{+} (\equiv W + \delta W)$ and $W_{-} (\equiv W - \delta W)$,
respectively, from which the differences from the standard abundance 
($A$) were derived as $\delta _{W+} (>0)$ and $\delta _{W-} (<0)$.

Figure~5 (C) and figure~6 (O) graphically show the resulting equivalent widths 
($W$; with $\pm \delta W$ as error bars), non-LTE corrections ($\Delta$), non-LTE abundances 
($A^{\rm N}$; with $\delta _{W+}$ and $\delta _{W-}$ as error bars), and abundance 
variations in response to parameter changes ($\delta_{T\pm}$, $\delta_{g\pm}$, 
and $\delta_{\xi\pm}$), as functions of $T_{\rm eff}$. We can recognize from these 
figures that (a) non-LTE corrections are very small ($\ltsim$ a few hundredths dex)
and practically negligible, (b) abundance errors due to parameter uncertainties 
(essentially due to ambiguities in $T_{\rm eff}$) amount to $\sim 0.1$~dex for C 
and $\ltsim$~0.2~dex for O (i.e., O abundances are subject to comparatively 
larger errors than C abundances), and (c) S/N-related abundance errors
are not so significant except for $A^{\rm N}$(O) of lower-$T_{\rm eff}$ stars
(6000~K~$\ltsim T_{\rm eff} \ltsim 6500$~K) where the O~{\sc i} line becomes
appreciably weak ($W \sim 10$~m$\rm\AA$ or less).

Since the present abundance study is based on a differential analysis,
the results should not be significantly affected by uncertainties in 
line-formation or model atmospheres such as 3D hydrodynamical effect 
(e.g., Amarsi et al. 2016, who showed that the 3D correction
for the O~{\sc i} 6158.2 line is on the order of $\sim$~0.1~dex and almost 
$T_{\rm eff}$-independent) or treatment of convection (such as with/without 
overshooting, resulting in abundance differences of $\sim$~0.1~dex; cf. Paper~I) 
which we do not explicitly take into consideration. In particular, thanks to 
the fact that C and O abundances are derived from similar high-excitation lines 
of neutral species, the resulting [C/O] ratio (the main purpose) must be
practically insensitive to such modeling details.

\subsection{Differential abundances relative to the Sun}

The differential abundances ([C/H], [O/H], and [Fe/H]) to be discussed in section~5
are computed as [C/H]~$\equiv$ $A^{\rm N}_{5380}$(C) $-$  $A^{\rm N}_{5380\odot}$(C),
[O/H]~$\equiv$ $A^{\rm N}_{6158}$(O) $-$  $A^{\rm N}_{6158\odot}$(O),
[Fe/H]$_{53} \equiv$ $A^{\rm L}_{5375-90}$(Fe) $-$  $A^{\rm L}_{5375-90\;\odot}$(Fe), and
[Fe/H]$_{61} \equiv$ $A^{\rm L}_{6150-67}$(Fe) $-$  $A^{\rm L}_{6150-67\;\odot}$(Fe),
where the adopted reference solar abundances of C, O, and Fe are
$A^{\rm N}_{5380\odot}$(C) = 8.43, $A^{\rm N}_{6158\odot}$(O) = {\it 8.81},\footnote{
Only for this O~{\sc i} 6158 line, we did not adopt the $A^{\rm N}$(O) solution 
of 8.68 derived from the 6150--6167~$\rm\AA$ region fitting of the Moon spectrum 
in this study but used the value (8.81) derived in Paper~I from the 
6156.4--6158.6~$\rm\AA$ fitting. This is because the strength of this O~{\sc i} 6158 
line becomes considerably weak ($W \sim$~2--3~m$\rm\AA$) at the solar temperature 
and the $A$(O) solution resulting from fitting is appreciably sensitive to a 
slight difference in the vertical offset (i.e., constant $C$ defined in Takeda 1995),
which is affected by the difference in the adopted spectral range and in the
treatment of 6157.4~$\rm\AA$ feature. 
In the present case, solution from the fitting in the narrower wavelength region 
as done in Paper~I is more reliable than that from the wider-region fitting 
such as adopted in this study. See appendix~1, where the reasoning of 
this choice is described in more detail.}
$A^{\rm L}_{5375-90\;\odot}$(Fe) = 7.35, and $A^{\rm L}_{6150-67\;\odot}$(Fe) = 7.60
(cf. tableE.txt). The results for Procyon turned out [C/H] = $-0.01$, [O/H] = +0.04,
[Fe/H]$_{53}$ = +0.05, and [Fe/H]$_{61}$ = $-0.07$; i.e. almost equal to the solar abundances,
which is consistent with the already known results (see, e.g., Sect.~IVc in 
Takeda et al. 2008).

\section{Results and discussion}

\subsection{Rotation and metallicity}

We first examine the rotational velocity ($v_{\rm e}\sin i$) and metallicity ([Fe/H]). 
resulting from the fitting analysis at 5375--5390~$\rm\AA$ and 6150--6167~$\rm\AA$. 
The $v_{\rm e}\sin i$ values derived from these two wavelength regions (ranging from
$\sim 10$~km~s$^{-1}$ to $\sim 110$~km~s$^{-1}$) are
in good agreement with each other (figure~7a), and show a decreasing trend with
a lowering of $T_{\rm eff}$ (figure~7b). While this is a trend well-known for 
field stars (i.e., manifest dropdown as the spectral type becomes 
later from F to G; see Fig.~18.21 in Gray 2005) and also seen in Hyades stars
(see Fig.~15a in Paper~I), the rotational velocities of Pleiades stars
tend to be comparatively higher reflecting their younger ages.
  
Regarding the metallicity, [Fe/H]$_{53}$ and [Fe/H]$_{61}$ mostly agree with each other 
except for several stars showing appreciable deviation by $\sim$~0.2--0.3~dex (figure~7c).
A close inspection revealed that discrepancies are seen only for rapidly rotating stars 
($v_{\rm e}\sin i \gtsim$~50~km~s$^{-1}$) while a consistency is confirmed 
for slower rotators of $v_{\rm e}\sin i \ltsim 50$~km~s$^{-1}$ (cf. figure~7d). 
According to figure~7d, [Fe/H]$_{61}$ appears to be somewhat $v_{\rm e}\sin i$-dependent
while such a trend is not seen for [Fe/H]$_{53}$, which makes us suspect
that the former is less reliable than the latter.
This difference may be related to the fact that more Fe lines of appreciable strengths are 
available in the 5375--5390~$\rm\AA$ region (figure~3) than in the 6150--6167~$\rm\AA$ 
region (figure~4), 

Accordingly, we adopted the values derived from the 5375--5390~$\rm\AA$ region
($v_{\rm e}\sin i_{53}$ and [Fe/H]$_{53}$) as the final results in table~2.
Figures~2d ($v_{\rm e}\sin i$) and 2e ([Fe/H]) show the comparison of these 
with Funayama et al.'s (2009) results, where we can see a fairly good
agreement for $v_{\rm e}\sin i$ while some (though not so serious) discrepancy 
up to $\ltsim 0.2$~dex is seen for [Fe/H].

\subsection{C and O abundances}

We now discuss the trends of C and O abundances as well as C/O ratios 
of the Pleiades stars. 
The finally resulting values of [Fe/H], [C/H], [O/H], and [C/O] 
for the 32 program stars are plotted against $T_{\rm eff}$ in figure~8,
where combined uncertainties due to parameter ambiguities and data noises 
(cf. subsection 4.2) are shown by error bars.
The following characteristics are read from these figures:\\
--- No systematic dependence upon $T_{\rm eff}$ is observed in these results
and thus almost homogeneous.\\
--- However, while [Fe/H] appears to be nearly solar,
we can see a trend of subsolar [C/H]~($<0$) and supersolar [O/H]~($>0$), 
resulting in a manifest tendency of subsolar [C/O]~$<0$.\\
--- This can be quantitatively confirmed. Since abundances from rapid 
rotators tend to be less reliable (as seen for the case of [Fe/H]$_{61}$ 
in subsection~5.1), we confine ourselves only to sharp-line stars of 
$v_{\rm e}\sin i < 50$~km~s$^{-1}$ (filled symbols in figure~8). Then, we have 
$\langle$[Fe/H]$\rangle$ = $0.00$ ($\sigma = 0.08$),
$\langle$[C/H]$\rangle$ = $-0.12$ ($\sigma = 0.06$),
$\langle$[O/H]$\rangle$ = $+0.08$ ($\sigma = 0.16$), and
$\langle$[C/O]$\rangle$ = $-0.20$ ($\sigma = 0.14$), 
which further result in the Fe-scaled abundance ratios of 
$\langle$[C/Fe]$\rangle$ = $-0.12$ ($\sigma = 0.10$) and 
$\langle$[O/Fe]$\rangle$ = $+0.08$ ($\sigma = 0.21$).

Consequently, regarding the C and O abundances of Pleiades F--G stars 
relative to the Sun, we conclude that [C/H] is underabundant by $\sim -0.1$~dex,
[O/H] is overabundant by $\sim +0.1$~dex, and thus [C/O] is underabundant
by $\sim -0.2$~dex, each holding with a dispersion of $\ltsim$~0.1--0.2~dex.

\subsection{[C/O] ratio of the Pleiades cluster and its implication}

We thus found that [C/O] in the Pleiades cluster is apparently subsolar by 
$\sim -0.2$~dex as a result of subsolar C and supersolar O.  
This means that the trend of negative [C/O] suspected in Hyades stars
([C/O] $\sim -0.1$~dex; cf. Sect.~6.2 in Paper~I) is more accentuated
in Pleiades stars, which makes contrast with the case of field solar-type stars
of near-solar metallicity ([Fe/H]~$\sim 0$) where [C/O]~$\sim 0$ holds 
on the average (cf. Fig.~6 in Takeda \& Honda 2005).  
In order to clarify this fact, we recalculated the average values of 
$\langle$[Fe/H]$\rangle$, $\langle$[C/Fe]$\rangle$, $\langle$[O/Fe]$\rangle$, 
and $\langle$[C/O]$\rangle$ for Hyades stars (here, only those with 
$T_{\rm eff} > 5800$~K were used to make the sample as similar as possible)\footnote{
The mean results for Hyades stars with $T_{\rm eff} > 5800$~K are: 
$\langle$[Fe/H]$\rangle$ = $+0.06$ ($\sigma = 0.10$),
$\langle$[C/H]$\rangle$ = $+0.12$ ($\sigma = 0.09$),
$\langle$[O/H]$\rangle$ = $+0.21$ ($\sigma = 0.20$), and
$\langle$[C/O]$\rangle$ = $-0.08$ ($\sigma = 0.19$).
Accordingly, [X/Fe] ratios for Hyades are
$\langle$[C/Fe]$\rangle$ = $+0.06$ ($\sigma = 0.09$) and 
$\langle$[O/Fe]$\rangle$ = $+0.14$ ($\sigma = 0.23$).
} 
as we have done for the Pleiades stars in subsection~5.2.
These mean abundances obtained for both Hyades and Pleiades are plotted by
large symbols in figure~9, where the corresponding values of field stars 
(the results were derived by using C~{\sc i} 5380 and O~{\sc i} 6158 lines 
while restricting to only F--G stars of $T_{\rm eff} > 5800$~K) are also shown. 
We can confirm by inspecting figure~9c that the $\langle$[C/O]$\rangle$ values 
for Hyades and Pleiades stars locate at the lower envelope of the 
[C/O] vs. [Fe/H] distribution of field stars, which suggests that 
the C/O ratios of these two open clusters rather deviate from the 
mean trend exhibited by the latter.
Why do the C/O ratios of cluster stars ([C/O]~$\sim -0.2$ to $-0.1$) 
are so different from those of F--G stars (of near-solar metallicity) 
and the Sun ([C/O]~$\sim 0$)? 
Does it have anything to do with the clear distinction in stellar ages
between the former ($\sim$~120--130~Myr for Pleiades and $\sim$~625~Myr 
for Hyades; cf. section~1) and the latter ($\sim$~1--10~Gyr; 
cf. Fig.~14 of Takeda 2007) ?

Here, it is important to note that the result we derived for Pleiades F--G 
stars is quite similar to that recently reported for B-type stars. 
That is, Nieva and Przybilla (2012) determined the light-element 
abundances for 20 B-type stars (16000~K~$\ltsim T_{\rm eff} \ltsim 33000$~K, 
$6 M_{\odot} \ltsim M \ltsim 19 M_{\odot}$) and concluded that the 
mean C and O abundances relative to Asplund et al.'s (2009) solar 
abundances are $\langle$[C/H]$\rangle = -0.10$ and 
$\langle$[O/H]$\rangle = +0.07$ (see Table~9 or Fig.~12 of their paper); 
i.e., subsolar [C/H] as well as supersolar [O/H] by $\sim 0.1$~dex, 
resulting in [C/O]~$\sim -0.2$~dex, which is just as we have obtained 
for the Pleiades stars.  Considering that these B-type stars around
$M \sim 10 M_{\odot}$ have typical ages of several tens Myr (e.g., 
Georgy et al. 2013), this result of subsolar [C/O] appears to be
the common characteristics seen for young stars born within 
$\ltsim 10^{8}$~yr. 

Therefore, the same discussion as developed by Nieva and Przybilla (2012) 
regarding the Sun and B-type stars may hold for the present case.
Their important suggestion was that the Sun was born in comparatively 
inner/metal-rich region at the galactocentric radius of $R_{\rm G} \sim$~5--6~kpc 
and has migrated outward to the current position of $R_{\rm G} \sim$~8~kpc, 
which naturally explains the paradox why the light-element abundances of 
such young stars are nearly the same (or even slightly underabundant) 
compared to the Sun (age of $4.6\times 10^{9}$~yr) despite that 
chemical evolution must have enriched the Galactic gas during the passage
of time. As they pointed out, this hypothesis leads to a reasonable account 
for the larger C-to-O abundance ratio of the Sun 
compared to those of young stars in the solar neighborhood, because 
C/O in the Galactic disk tends to decline with an increase of 
$R_{\rm G}$ as indicated from recombination-line observations of 
H~{\sc ii} regions (Esteban et al. 2005) and supported by some 
theoretical models of chemical evolution (Carigi et al. 2005).

Consequently, according to this scenario, our observational fact 
that the [C/O] values in the Pleiades F--G stars are by $\sim -0.2$~dex 
lower than the Sun may be interpreted as due to the difference of 
C/O ratios in two kinds of interstellar gases:
(1) which recently formed young stars near to the current Sun 
($R_{\rm G} \sim$~8~kpc, age of $\ltsim 10^{8}$~yr) and (2) which 
formed the Sun in the inner region ($R_{\rm G} \sim$~5--6~kpc) 
long before ($4.6\times 10^{9}$~yr ago).
We should remark, however, that the [C/O] values of field  
stars of near-solar metallicity ([Fe/H]~$\sim 0$) studied by 
Takeda and Honda (2005) evenly distribute around $\sim 0$ (figure~9c), 
which means that the Sun is not exceptional compared to nearby solar-type stars 
in terms of the C/O ratio. So, if our Sun really has significatly changed
its orbit since its birth, such a migration may not necessarily be 
a rare phenomenon for stars in the Galactic disk. 

In summary, while we found based on our C and O abundance study 
of 32 Pleiades stars that the [C/O] ratios of these recently 
formed young stars are by $\sim -0.2$~dex lower than those of 
the Sun and similarly aged solar-metalicity field stars,
this difference between these two age groups is presumably due to
(not the direct time-evolution effect but) the orbit migration mechanism 
which Galactic stars may undergo in a long time.
Given this scenario, detecting C-rich stars with $N$(C)/$N$(O)$>0.8$ 
(mentioned in section~1) would be hardly possible among young stars
in the solar neighborhood, while such a possibility may be higher
for old metal-rich stars which were born near to the Galactic bulge. 

\bigskip

Special thanks are due to H. Funayama for his substantial contribution
in obtaining the observational data for 25 Pleiades stars 
in the 2006--2007 season, upon which this study heavily depends.

This research has made use of the SIMBAD database (operated by
CDS, Strasbourg, France) as well as the VALD database (operated 
at Uppsala University, the Institute of Astronomy RAS in Moscow, 
and the University of Vienna).

Data reduction was in part carried out by using the common-use data analysis 
computer system at the Astronomy Data Center (ADC) of the National Astronomical 
Observatory of Japan.

\appendix

\section{Solar oxygen abundance derived from the O~{\sc i} 6156--8 lines}

In deriving differential [O/H] (subsection~4.3), we intentionally adopted 
$A_{\odot}$(O) = 8.81 derived by Takeda and Honda (2005) as the reference 
solar oxygen abundance, instead of the slightly lower value of 8.68 
obtained in this study, since we believe that the former is more reliable
than the latter. This is related to the difficulty involved with O-abundance
determination from O~{\sc i} 6156--8 lines in the spectrum of the Sun,
where lines of other species are blended because of its comparatively 
low $T_{\rm eff}$. Actually, there are differences between these two
analyses, although they used the same model atmosphere and the same Moon 
spectrum.\\
--- First, Takeda and Honda (2005) carried out the fitting in
the limited 6156.4--6158.6~$\rm\AA$ region (only 2.2~$\rm\AA$ wide) while
we applied fittings to a much wider 6150--6167~$\rm\AA$ region in this study
since stars with considerably broad lines (up to 
$v_{\rm e}\sin i \sim 100$~km~s$^{-1}$) are involved (i.e., to ensure the
stability of solutions).\\
--- Second, although both studies were based on the same Kurucz and Bell's 
(1995) compilation for the atomic line data, Takeda and Honda (2005)
applied arbitrary chages in the $gf$ values of 4 Ti lines (cf. footnote~5 
therein) in order to reproduce the absorption feature at $\sim 6157.4$~$\rm\AA$,
while such modifications were not applied in this study where 
all the data given in Kurucz and Bell (1995) were employed without any 
change since such a $\sim 6157.4$~$\rm\AA$ feature is seen only for 
lower-$T_{\rm eff}$ stars ($T_{\rm eff} \ltsim 6000$~K; cf. Fig.~1 
in Takeda \& Honda 2005) and irrelevant for most of our sample stars.

The resulting $A_{\odot}$(O) solutions for these two cases show a difference 
by 0.13~dex (8.81 and 8.68) because of the different wavelength span 
and different treatment for the $\sim 6157.4$~$\rm\AA$ feature (figure~10a).
However, we have a good reason to believe that the former solution (8.81) 
is more reliable than the latter (8.68), as manifested from the simulated strength 
of the strongest 6158.2~$\rm\AA$ line (which is considered to be practically 
blend-free and most reliable among the three O~{\sc i} features, in contrast to 
those at 6156.0 and 6156.8~$\rm\AA$ being appreciably contaminated by lines of 
other species) in comparison with Kurucz et al.'s (1984) solar flux spectrum of 
very high resolving power (figure~10b). (Besides, if we adopt 8.68, we have
[O/H]$_{\rm Procyon} = +0.17$, which is unacceptably too high for this star 
known to have near-solar composition.\footnote{
The oxygen abundance of Procyon (= HR~2943 = HD~61421 = HIP~37279) 
relative to the Sun has already been determined in several published 
studies using various lines of different reliability. 
For example, according to Steffen's (1985) analysis, 
[O/H]$_{\rm Procyon}$ is $-0.17$ (from O~{\sc i} 6156.8; unreliable 
because of the blending in the solar spectrum as mentioned above), 
$-0.01$ (from O~{\sc i} 6158.2; most credible), 
and $-0.05$ ([O~{\sc i}] 6300.3). Edvardsson et al. (1993) derived
[O/H]$_{\rm Procyon} = -0.05$ (O~{\sc i} 6158.2, O~{\sc i} 7771--5),
while [O/H]$_{\rm Procyon}= +0.02$ (from [O~{\sc i}] 6300, O~{\sc i} 7771--5) 
was reported in Allende Prieto et al.'s (2004) S$^4$N project 
summarized at $\langle$http://hebe.as.utexas.edu/s4n/$\rangle$.
Further, Takeda and Honda (2005) derived [O/H]$_{\rm Procyon} = -0.04$ 
(from [O~{\sc i}]~6300), $+0.06$ (from O~{\sc i}~6157--8; finally 
adopted by them), and +0.07 (from O~{\sc i} 7771--5). 
These results indicate that the photospheric O abundances of Procyon 
and Sun are quite similar to a degree of several hundredths dex.}) 
It should also be remarked that a consistency is accomplished 
by this choice of using Takeda and Honda's (2005) $A_{\odot}$(O) = 8.81 
with regard to the comparison of figure~9, where the mean values of 
[O/Fe] and [C/O] for the Pleiades stars are compared 
with those of field F--G stars derived in Takeda and Honda (2005). 

Regarding the lines of other species contaminating the relevant 
O~{\sc i} lines in the solar spectrum, the weakest O~{\sc i}~6156.0~$\rm\AA$ line 
is severely blended with the Ca~{\sc i}~6156.023 line which we took into
consideration (cf. table~3). 
Meanwhile, the O~{\sc i}~6156.8 line (medium strength) is evidently blended 
with some other line (cf. figure~10b) which is not included in Kurucz and Bell's (1995)
compilation. This unknown blend may be due to Fe~{\sc i}, since the VALD database
contains Fe~{\sc i} 6156.804 ($\chi_{\rm low}$ = 4.956~eV, $\log gf = -1.495$),
though this $\log gf$ must be erroneously too large  as it predicts too strong 
contribution [$W_{\odot}~\sim 13$~m$\rm\AA$ for $A_{\odot}$(Fe) = 7.50] compared 
to the required amount of only $\sim 2$~m$\rm\AA$.
Finally, although Pereira, Asplund, and Kiselman (2009) suspected that the 
strongest O~{\sc i} 6158.2 line is blended by weak CN lines, this effect 
(if any exists) would not be so significant, because Bertran de Lis et al. (2015) 
reported that the oxygen abundances derived from O~{\sc i} 6158.2 and 
[O~{\sc i}] 6300.30 lines agree well with each other.
In any event, as the strengths of such Fe~{\sc i} as well as CN lines progressively 
decrease with an increase in $T_{\rm eff}$, their blending effect is negligible 
for most of our Pleiades stars which have $T_{\rm eff} \gtsim 6000$~K.

\section{Lithium abundances of 7 Pleiades stars}

We originally wished to organize this study on Pleiades stars in close analogy 
with Paper~I, where not only C and O abundances but also Li abundances 
of Hyades F--G stars were determined. Unfortunately, given that we 
primarily relied on the available archived data for many of our sample stars 
(25 out of 32), Li abundances for those stars could not be derived as the 
Li~{\sc i} 6708 line was outside the wavelength range of their spectra (cf. table~1).
Yet, since this Li line was included in the spectra of 7 stars (i.e., No. 26--32
in table~1) newly observed by ourselves in the 2015--2016 season, we could 
determine their Li abundances, which we briefly report here for reference. 

The procedure for Li abundance determination is essentially the same as
described in Takeda and Kawanomoto (2005) (also adopted in Paper~I), 
which may be consulted for more details.
The best-fit theoretical spectrum with the observed data in the 6702--6712~$\rm\AA$ 
region is shown in figure~10 for each star. The resulting equivalent width 
of the Li~{\sc i} 6708 doublet (in m$\rm\AA$), non-LTE abundance (in the usual 
normalization of $A$(H)=12), and non-LTE correction (in dex) are ($W$, $A^{\rm N}$(Li), 
$\Delta$) = (81.3, 3.07, 0.00), (168.2, 3.50, +0.07), 
(137.7, 3.10, +0.10), (142.5, 3.34, +0.05), (55.6, 3.00, $-0.03$), 
(144.4, 3.30, +0.06), and (123.6, 3.24, +0.04), for
HD~23269 (No.~26), HD~23386 (No.~27), MSK~65 (No.~28), BD~+22~548 (No.~29), 
HD~23513 (No.~30), HD~282967 (No.~31), and HD~283067 (No.~32), respectively.

If these $A^{\rm N}$(Li) values of 7 stars (ranging at 
5800~K~$\ltsim T_{\rm eff} \ltsim 6600$~K, overlapping the
Hyades Li-dip of $T_{\rm eff} \sim$~6200--6800~K) are overplotted
on the $A^{\rm N}$(Li) vs. $T_{\rm eff}$ relation of Hyades stars
presented in Fig.~14e of Paper~I, we can confirm that the primordial
Li abundance ($\sim 3.3$) is nearly preserved in the photosphere of 
Pleiades stars at $T_{\rm eff} \sim$~5800--6600~K with little sign of depletion,
in marked contrast with the case of Hyades stars at the Li chasm. 
This is consistent with the results of previous studies
(see, e.g., Fig.~1 of Boesgaard, Armengaud, \& King 2003).

\clearpage

\onecolumn

\setcounter{table}{0}
\begin{table}[h]
\small
\caption{Basic observational data of the program stars.}
\begin{center}
\begin{tabular}
{rccrrcrrl}\hline \hline
No. & Star & HII\# & $V$ & $B-V$ & Instr. & $t_{\rm exp}^{\rm total}$ & S/N & Obs. date \\
(1) & (2)    & (3)   & (4) & (5)   & (6)    & (7)                       & (8) & (9)       \\
\hline
   1  &  V1038 Tau  &      314&  10.66 &  0.66 &  G1  &  420  &   60 &  2007 Jan 13, 14, 15         \\
   2  &  HD 23289   &      470&   8.86 &  0.40 &  G1  &   81  &  120 &  2006 Jan 25                 \\
   3  &  MSK 39     &      489&  10.41 &  0.63 &  G1  &  510  &   80 &  2006 Jan 30, 2007 Jan 11, 12\\
   4  &  HD 23325   &      531&   8.58 &  0.34 &  H   &   65  &  150 &  2006 Jan 11                 \\
   5  &  HD 23375   &      697&   8.59 &  0.35 &  G1  &   30  &   40 &  2007 Jan 15                 \\
   6  &  V855 Tau   &      727&   9.71 &  0.56 &  G1  &   44  &   60 &  2006 Jan 25                 \\
   7  &  HD 23585   &     1284&   8.37 &  0.29 &  H   &   30  &  150 &  2006 Jan 11                 \\
   8  &  HD 23584   &     1309&   9.46 &  0.47 &  H   &   90  &  100 &  2006 Jan 15                 \\
   9  &  HD 282972  &     1794&  10.36 &  0.66 &  G1  &  180  &   70 &  2006 Jan 28                 \\
  10  &  BD +23 551 &     1797&  10.13 &  0.56 &  G1  &  132  &  100 &  2006 Jan 27                 \\
  11  &  BD +22 574 &     2506&  10.27 &  0.60 &  G1  &  190  &  100 &  2006 Jan 29                 \\
  12  &  HD 24194   &     3179&  10.05 &  0.57 &  H   &  240  &  120 &  2006 Dec 6                  \\
  13  &  BD +23 472 &     3301&   9.54 &  0.51 &  H   &   30  &   40 &  2006 Dec 10                 \\
  14  &  HD 22887   &     3309&   9.18 &  0.45 &  H   &   60  &   80 &  2006 Jan 12                 \\
  15  &  HD 22977   &     3310&   9.11 &  0.47 &  H   &   90  &  120 &  2006 Jan 12, Dec 10         \\
  16  &  HD 23312   &     3314&   9.49 &  0.49 &  H   &  180  &  130 &  2006 Jan 12, 15             \\
  17  &  HD 23488   &     3319&   8.73 &  0.37 &  H   &   75  &  150 &  2006 Jan 11                 \\
  18  &  HD 23975   &3326/5026&   9.64 &  0.52 &  H   &  150  &  160 &  2006 Dec 10                 \\
  19  &  BD +21 503 &  $\cdots$  &  10.35 &  0.62 &  G1  &  480  &   70 &  2007 Jan 10, 11             \\
  20  &  BD +22 521 &  $\cdots$  &  10.07 &  0.54 &  H   &  240  &   80 &  2006 Dec 11. 12             \\
  21  &  BD +23 486 &  $\cdots$  &  10.34 &  0.60 &  G1  &  300  &   80 &  2007 Jan 15, 18             \\
  22  &  BD +25 591 &  $\cdots$  &  10.06 &  0.56 &  H   &  180  &  100 &  2006 Dec 10                 \\
  23  &  BD +25 610 &     3442&   9.99 &  0.59 &  G1  &  360  &   90 &  2007 Jan 9, 10              \\
  24  &  HD 23598   &  $\cdots$  &   9.83 &  0.54 &  H   &   60  &   50 &  2006 Dec 6                  \\
  25  &  HD 24655   &     3332&   9.07 &  0.45 &  H   &   85  &  100 &  2006 Jan 12, Dec 9          \\
  26  &  HD 23269   &      405&   9.84 &  0.54 &  G2  &  360  &  120 &  2016 Jan 14                 \\
  27  &  HD 23386   &      739&   9.57 &  0.62 &  G2  &  180  &  100 &  2015 Dec 18                 \\
  28  &  MSK 65     &      761&  10.56 &  0.71 &  G2  &  300  &  120 &  2016 Mar 16, 22, 24, 25     \\
  29  &  BD +22 548 &      923&  10.16 &  0.61 &  G2  &  270  &  110 &  2016 Feb 2                  \\
  30  &  HD 23513   &     1139&   9.37 &  0.48 &  G2  &  240  &  150 &  2015 Nov 30                 \\
  31  &  HD 282967  &     1514&  10.48 &  0.64 &  G2  &  360  &   90 &  2016 Feb 15                 \\
  32  &  HD 283067  &     2786&  10.31 &  0.60 &  G2  &  300  &  110 &  2016 Mar 1, 15              \\

\hline
\end{tabular}
\end{center}
\footnotesize
(1) Object number (arbitrarily assigned). (2) Star name. (3) Hertzsprung's (1947) HII number.
(4) Observed $V$ magnitude. (5) Observed $B-V$ color. (6) Key to the used instrument
(G1 $\cdots$ GAOES data in 2006--2007 covering 4870--6670~$\rm\AA$,
 G2 $\cdots$ GAOES data in 2015--2016 covering 4950--6800~$\rm\AA$,
 H $\cdots$ HIDES data in 2006 covering 5340--6580~$\rm\AA$).
(7) Total exposure time (in minutes). (8) Signal-to-noise ratio directly measured around 
$\sim$~5370--5390~$\rm\AA$. (9) Dates of observations.
  
\end{table}

\setcounter{table}{1}
\begin{table}[h]
\small
\caption{Stellar parameters and the results of abundance analysis.}
\begin{center}
\begin{tabular}
{rc ccc r c rrc rrc}\hline \hline
No. & Object & $T_{\rm eff}$ & $\log g$ & $\xi$ & $v_{\rm e}\sin i$ & [Fe/H] & 
$W^{\rm C}_{5380}$ & $\Delta^{\rm C}_{5380}$ & [C/H] & $W^{\rm O}_{6158}$ & $\Delta^{\rm O}_{6158}$ & [O/H] \\
(1) & (2) & (3) & (4) & (5) & (6) & (7) & (8) & (9) & (10) & (11) & (12) & (13) \\
\hline
  1 &  V1038 Tau  & 5982 & 4.50 & 1.2 & 36 &+0.09& 21.2& $-$0.01& $-$0.03&  9.9&  0.00& +0.32\\
  2 &  HD 23289   & 6945 & 4.32 & 2.2 & 27 &+0.05& 37.9& $-$0.02& $-$0.14& 20.5& $-$0.02& $-$0.05\\
  3 &  MSK 39     & 6067 & 4.48 & 1.2 & 18 &+0.01& 23.6& $-$0.01& $-$0.02&  8.1&  0.00& +0.14\\
  4 &  HD 23325   & 7289 & 4.31 & 2.5 & 78 &+0.27& 31.3& $-$0.02& $-$0.35& 35.6& $-$0.04& +0.04\\
  5 &  HD 23375   & 7227 & 4.31 & 2.5 & 87 &$-$0.02& 30.9& $-$0.02& $-$0.34& 54.2& $-$0.04& +0.35\\
  6 &  V855 Tau   & 6266 & 4.44 & 1.5 & 51 &+0.01& 21.1& $-$0.01& $-$0.19&  7.3&  0.00& $-$0.07\\
  7 &  HD 23585   & 7629 & 4.31 & 2.9 &110 &+0.22& 54.6& $-$0.02& $-$0.10& 50.4& $-$0.05& +0.01\\
  8 &  HD 23584   & 6612 & 4.37 & 1.8 & 78 &$-$0.10& 30.3& $-$0.01& $-$0.15& 21.2& $-$0.01& +0.22\\
  9 &  HD 282972  & 5982 & 4.50 & 1.2 & 14 &+0.04& 18.3& $-$0.01& $-$0.11&  4.2&  0.00& $-$0.11\\
 10 &  BD +23 551 & 6266 & 4.44 & 1.5 & 21 &$-$0.02& 21.7& $-$0.01& $-$0.17&  8.7&  0.00& +0.02\\
 11 &  BD +22 574 & 6150 & 4.46 & 1.3 & 15 &+0.01& 23.9& $-$0.01& $-$0.06&  9.0&  0.00& +0.13\\
 12 &  HD 24194   & 6235 & 4.45 & 1.4 &  8 &$-$0.01& 23.0& $-$0.01& $-$0.12&  9.2&  0.00& +0.07\\
 13 &  BD +23 472 & 6446 & 4.40 & 1.7 & 43 &$-$0.09& 27.7& $-$0.01& $-$0.12& 15.2&  0.00& +0.16\\
 14 &  HD 22887   & 6700 & 4.35 & 1.9 & 92 &$-$0.14& 31.3& $-$0.02& $-$0.17& 25.5& $-$0.01& +0.26\\
 15 &  HD 22977   & 6612 & 4.37 & 1.8 & 57 &+0.04& 41.0& $-$0.02& +0.05& 22.2& $-$0.01& +0.25\\
 16 &  HD 23312   & 6527 & 4.38 & 1.7 & 12 &$-$0.14& 30.5& $-$0.01& $-$0.10& 17.4&  0.00& +0.17\\
 17 &  HD 23488   & 7108 & 4.31 & 2.3 & 19 &$-$0.06& 40.3& $-$0.02& $-$0.15& 31.7& $-$0.03& +0.09\\
 18 &  HD 23975   & 6407 & 4.41 & 1.6 & 23 &$-$0.17& 29.6& $-$0.01& $-$0.06& 14.7&  0.00& +0.17\\
 19 &  BD +21 503 & 6095 & 4.48 & 1.3 & 33 &$-$0.09& 22.5& $-$0.01& $-$0.06& 17.0&  0.00& +0.53\\
 20 &  BD +22 521 & 6333 & 4.42 & 1.5 & 27 &+0.16& 26.0& $-$0.01& $-$0.11&  7.8&  0.00& $-$0.10\\
 21 &  BD +23 486 & 6150 & 4.46 & 1.3 & 12 &+0.03& 21.6& $-$0.01& $-$0.12&  9.6&  0.00& +0.16\\
 22 &  BD +25 591 & 6266 & 4.44 & 1.5 & 35 &+0.02& 22.9& $-$0.01& $-$0.14& 13.1&  0.00& +0.23\\
 23 &  BD +25 610 & 6178 & 4.46 & 1.4 & 10 &$-$0.08& 20.4& $-$0.01& $-$0.16&  8.9&  0.00& +0.10\\
 24 &  HD 23598   & 6333 & 4.42 & 1.5 & 69 &+0.03& 31.6& $-$0.01& +0.01& 35.9&  0.00& +0.79\\
 25 &  HD 24655   & 6700 & 4.35 & 1.9 & 25 &$-$0.01& 38.4& $-$0.02& $-$0.04& 18.7& $-$0.01& +0.08\\
 26 &  HD 23269   & 6333 & 4.42 & 1.5 & 19 &+0.07& 26.8& $-$0.01& $-$0.09&  9.2&  0.00& $-$0.01\\
 27 &  HD 23386   & 6095 & 4.48 & 1.3 & 15 &+0.08& 19.9& $-$0.01& $-$0.13&  5.5&  0.00& $-$0.07\\
 28 &  MSK 65     & 5826 & 4.53 & 1.0 & 13 &$-$0.11& 12.3& $-$0.01& $-$0.24&  2.6&  0.00& $-$0.21\\
 29 &  BD +22 548 & 6123 & 4.47 & 1.3 & 19 &+0.12& 18.8& $-$0.01& $-$0.18&  4.6&  0.00& $-$0.18\\
 30 &  HD 23513   & 6569 & 4.37 & 1.8 & 33 &+0.01& 36.7& $-$0.02& $-$0.01& 18.8& $-$0.01& +0.18\\
 31 &  HD 282967  & 6039 & 4.49 & 1.2 & 16 &+0.05& 16.0& $-$0.01& $-$0.22&  6.3&  0.00& +0.04\\
 32 &  HD 283067  & 6150 & 4.46 & 1.3 & 25 &+0.02& 17.9& $-$0.01& $-$0.22&  8.4&  0.00& +0.09\\
\hline
\end{tabular}
\end{center}
\footnotesize
(1) Object number. (2) Star name. (3) Effective temperature (in K). (4) logarithm of surface gravity
(in cm~s$^{-2}$). (5) Microturbulence (in km~s$^{-1}$). (6) Projected rotational velocity
(in km~s$^{-1}$) derived from 5375--5390~$\rm\AA$ fitting. (7) Fe abundance (LTE) relative to the Sun 
(in dex) derived from 5375--5390~$\rm\AA$ fitting. (8) Equivalent width (in m$\rm\AA$) for 
C~{\sc i} 5380.325. (9) Non-LTE correction for C~{\sc i} 5380.325 (in dex). (10) Non-LTE carbon 
abundance relative to the Sun (in dex). (11) Equivalent width (in m$\rm\AA$) for O~{\sc i} 6158
comprising 3 components. (9) Non-LTE correction for O~{\sc i} 6158 (in dex). (10) Non-LTE oxygen 
abundance relative to the Sun (in dex). See subsection 4.3 for more details regarding the derivation 
of [Fe/H], [C/H], and [O/H].
\end{table}

\setcounter{table}{2}
\begin{table}[h]
\small
\caption{Atomic parameters of important lines relevant for spectrum fitting.}
\begin{center}
\begin{tabular}{ccrccccl}\hline\hline
Species & $\lambda$ & $\chi$ & $\log gf$ & Gammar & Gammas & Gammaw & Remark\\
(1) & (2) & (3) & (4) & (5) & (6) & (7) & (8) \\
\hline
\hline
\multicolumn{8}{c}{[5375--5390~$\rm\AA$ fitting]}\\
 Mn~{\sc i} & 5377.607 &  3.844 & $-$0.166 & 7.90 & $-$5.39 & $-$7.59 & \\
 C~{\sc i} & 5378.910 &  8.851 & $-$0.912 & 8.34 & $-$1.52 & $-$6.59 & minor contribution\\
 Fe~{\sc i} & 5379.573 &  3.695 & $-$1.514 & 7.85 & $-$6.13 & $-$7.58 & \\
 C~{\sc i} & 5380.227 &  8.851 & $-$1.066 & 8.11 & $-$1.86 & $-$6.58 & minor contribution\\
 C~{\sc i} & 5380.267 &  8.851 & $-$2.096 & 7.99 & $-$1.96 & $-$6.62 & negligible\\
 C~{\sc i} & 5380.325 &  7.685 & $-$1.616 & 8.69 & $-$5.65 & $-$7.37 & C~5380: most important \\
 Ti~{\sc ii} & 5381.022 &  1.566 & $-$1.970 & 8.15 & $-$6.59 & $-$7.85 & \\
 Fe~{\sc i} & 5382.263 &  5.669 & $-$0.252 & 8.33 & $-$4.39 & $-$7.39 & \\
 Fe~{\sc i} & 5383.368 &  4.312 &  +0.645 & 8.30 & $-$5.01 & $-$7.22 & \\
 Fe~{\sc i} & 5386.333 &  4.154 & $-$1.770 & 8.45 & $-$4.47 & $-$7.17 & \\
 Fe~{\sc i} & 5387.480 &  4.143 & $-$2.034 & 8.10 & $-$5.66 & $-$7.70 & \\
 C~{\sc i} & 5388.226 &  8.847 & $-$1.183 & 8.34 & $-$2.05 & $-$6.66 & minor contribution\\
 Fe~{\sc i} & 5389.478 &  4.415 & $-$0.410 & 8.32 & $-$4.81 & $-$7.16 & \\
\hline
\multicolumn{8}{c}{[6150--6167~$\rm\AA$ fitting]}\\
 Fe~{\sc i} & 6151.617 &  2.176 & $-$3.299 & 8.19 & $-$6.20 & $-$7.82 & \\
 Na~{\sc i} & 6154.226 &  2.102 & $-$1.560 & 7.85 & $-$4.39 & ($-$7.29) & \\
 Si~{\sc i} & 6155.134 &  5.619 & $-$0.400 & (7.77) & ($-$4.45) & ($-$7.05) & \\
 Si~{\sc i} & 6155.693 &  5.619 & $-$1.690 & (7.77) & ($-$4.45) & ($-$7.05) & \\
 O~{\sc i} & 6155.961 & 10.740 & $-$1.401 & 7.60 & $-$3.96 & ($-$7.23) & O~6156\\
 O~{\sc i} & 6155.971 & 10.740 & $-$1.051 & 7.61 & $-$3.96 & ($-$7.23) & O~6156\\
 O~{\sc i} & 6155.989 & 10.740 & $-$1.161 & 7.61 & $-$3.96 & ($-$7.23) & O~6156\\
 Ca~{\sc i}& 6156.023 &  2.521 & $-$2.200 & 7.49 & $-$4.69 & $-$7.50 & \\
 O~{\sc i} & 6156.737 & 10.740 & $-$1.521 & 7.61 & $-$3.96 & ($-$7.23) & O~6157\\
 O~{\sc i} & 6156.755 & 10.740 & $-$0.931 & 7.61 & $-$3.96 & ($-$7.23) & O~6157\\
 O~{\sc i} & 6156.778 & 10.740 & $-$0.731 & 7.62 & $-$3.96 & ($-$7.23) & O~6157\\
 Fe~{\sc i} & 6157.725 &  4.076 & $-$1.260 & 7.70 & $-$6.06 & $-$7.84 & \\
 O~{\sc i} & 6158.149 & 10.741 & $-$1.891 & 7.62 & $-$3.96 & ($-$7.23) & O~6158\\
 O~{\sc i} & 6158.172 & 10.741 & $-$1.031 & 7.62 & $-$3.96 & ($-$7.23) & O~6158\\
 O~{\sc i} & 6158.187 & 10.741 & $-$0.441 & 7.61 & $-$3.96 & ($-$7.23) & O~6158\\
 Na~{\sc i} & 6160.747 &  2.104 & $-$1.260 & 7.85 & $-$4.39 & ($-$7.29) & \\
 Ca~{\sc i} & 6161.297 &  2.523 & $-$1.020 & 7.49 & $-$4.69 & $-$7.50 & \\
 Ca~{\sc i} & 6162.173 &  1.899 &  +0.100 & 7.82 & $-$5.07 & $-$7.59 & \\
 Ni~{\sc i} & 6163.418 &  4.105 & $-$0.682 & 8.31 & $-$4.06 & $-$7.76 & \\
 Ca~{\sc i} & 6163.755 &  2.521 & $-$1.020 & 7.48 & $-$4.69 & $-$7.50 & \\
 Fe~{\sc i} & 6165.361 &  4.143 & $-$1.550 & 7.94 & $-$6.16 & $-$7.83 & \\
 Ca~{\sc i} & 6166.439 &  2.521 & $-$0.900 & 7.48 & $-$4.69 & $-$7.50 & \\
\hline
\end{tabular}
\end{center}
\footnotesize 
(1) Line species. (2) Air wavelength (in $\rm\AA$). (3) Lower excitation potential 
(in eV). (4) Logarithm of $g_{i}$ (statistical weight of the lower level) times 
$f_{ij}$ (absorption oscillator strength). (5) Logarithm of the radiation damping 
width in unit of s$^{-1}$ [$\log\gamma_{\rm rad}$] (6) Logarithm of the Stark 
damping width (s$^{-1}$) per electron density (cm$^{-3}$) at $10^{4}$ K 
[$\log(\gamma_{\rm e}/N_{\rm e})$]. (7) Logarithm of the van der Waals damping 
width (s$^{-1}$) per hydrogen density (cm$^{-3}$) at $10^{4}$ K 
[$\log(\gamma_{\rm w}/N_{\rm H})$]. (8) Miscellaneous remarks. 
The data in the 5375--5390~$\rm\AA$ region were adopted from 
the VALD database, while those in the 6150--6167~$\rm\AA$ region were
taken from the compilation of Kurucz and Bell (1995) in order to maintain
consistency with Paper~I. Parenthesized values of the damping parameters are those
computed according to the default treatment of Kurucz's (1993) WIDTH9 program 
(cf. Leusin, Topil'skaya 1987), since they are not given in the original database.
\end{table}

\clearpage

\begin{figure}
  \begin{center}
    \FigureFile(130mm,130mm){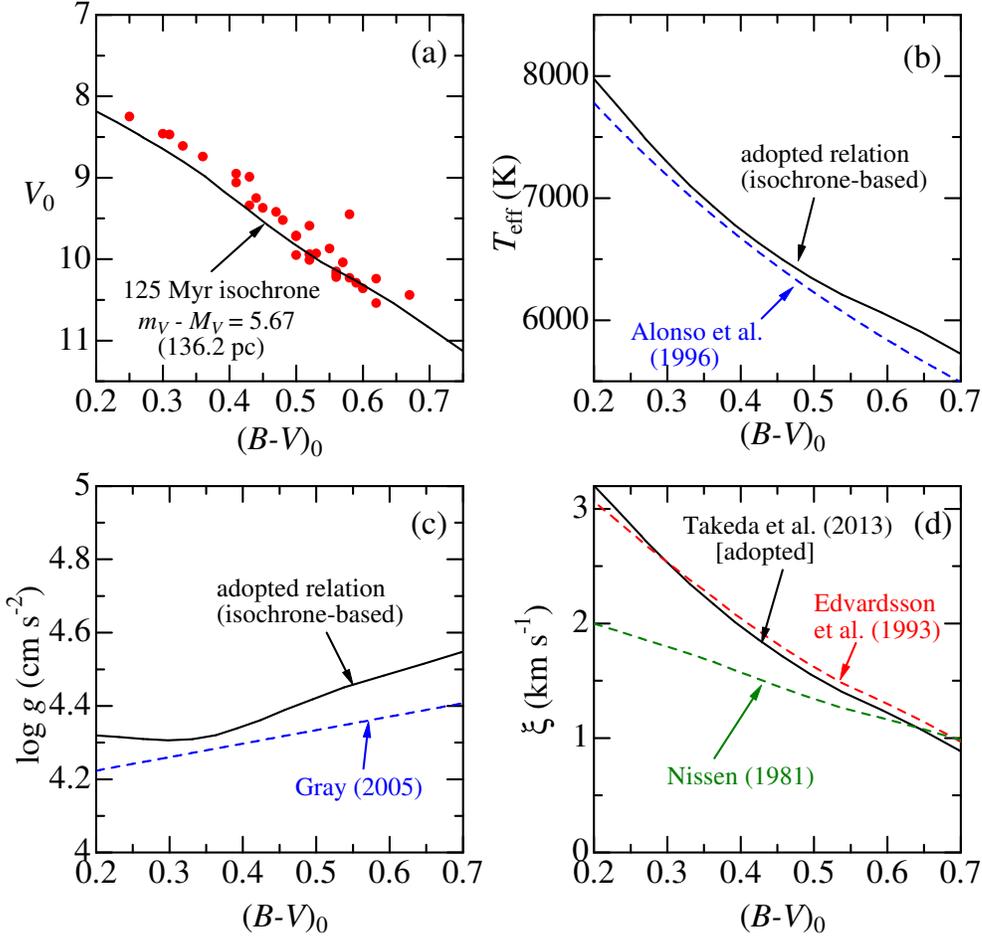}
  \end{center}
\caption{(a) Solid lines: Theoretical PARSEC isochrone (Bressan et al. 2012, 2013) 
of solar-metallicity corresponding to the age of 125~Myr, which was converted 
from the absolute to apparent magnitude scale by using the Pleiades distance 
of 136.2 pc (Melis et al. 2014). Symbols: $V_{0}$ vs. $(B-V)_{0}$ diagram
of the program stars, where the interstellar reddening effect
was corrected by assuming $E_{B-V} = A_{V}/3.1 = 0.03$. (b) Isochrone-based
$T_{\rm eff}$ vs. $(B-V)_{0}$ relation adopted for the Pleiades stars (solid line)
compared with Alonso et al.'s (1996) empirical calibration for the case of 
[Fe/H] = 0 (dashed line). (c) Isochrone-based $\log g$ vs. $(B-V)_{0}$ 
relation adopted for the Pleiades stars (solid lines), compared with Gray's (2005) 
empirical relation for dwarfs ($\log g = 4.15 + 0.368 (B-V)_{0}$; dashed line). 
(d) Adopted $\xi$ vs. $(B-V)_{0}$ relation corresponding to the empirical 
$\xi (T_{\rm eff}, \log g)$ formula derived in Paper~I (solid line),
compared with those of Edvardsson et al. (1993) and Nissen (1981) (dashed lines).
}
\end{figure}

\begin{figure}
  \begin{center}
    \FigureFile(130mm,130mm){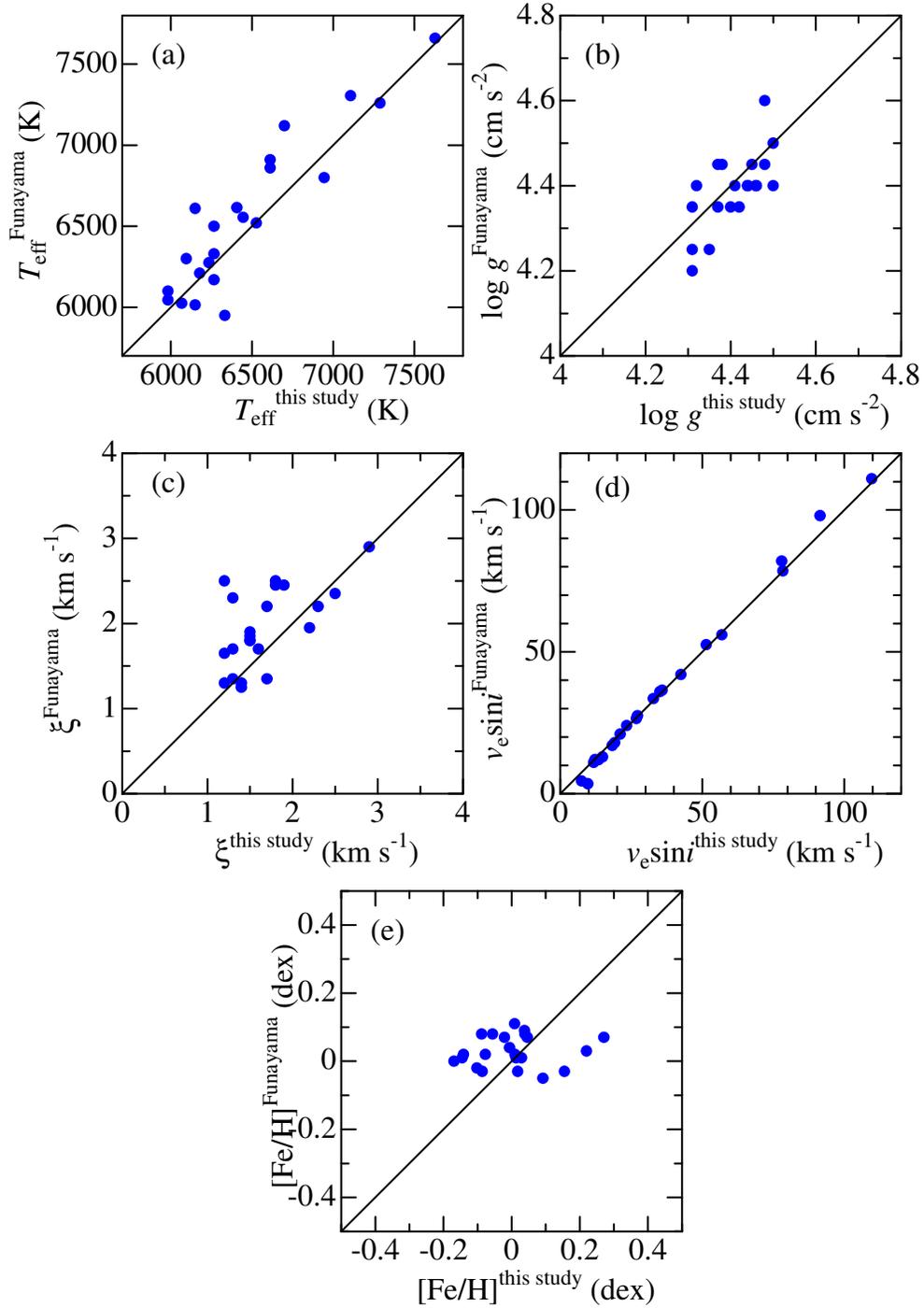}
  \end{center}
\caption{
Comparison of the parameters adopted in this study (as given in table~2)
with those derived spectroscopically by Funayama et al. (2009).
(a) $T_{\rm eff}$, (b) $\log g$, (c) $\xi$, (d) $v_{\rm e}\sin i$,
and (e) [Fe/H].
}
\end{figure}

\begin{figure}
  \begin{center}
    \FigureFile(150mm,200mm){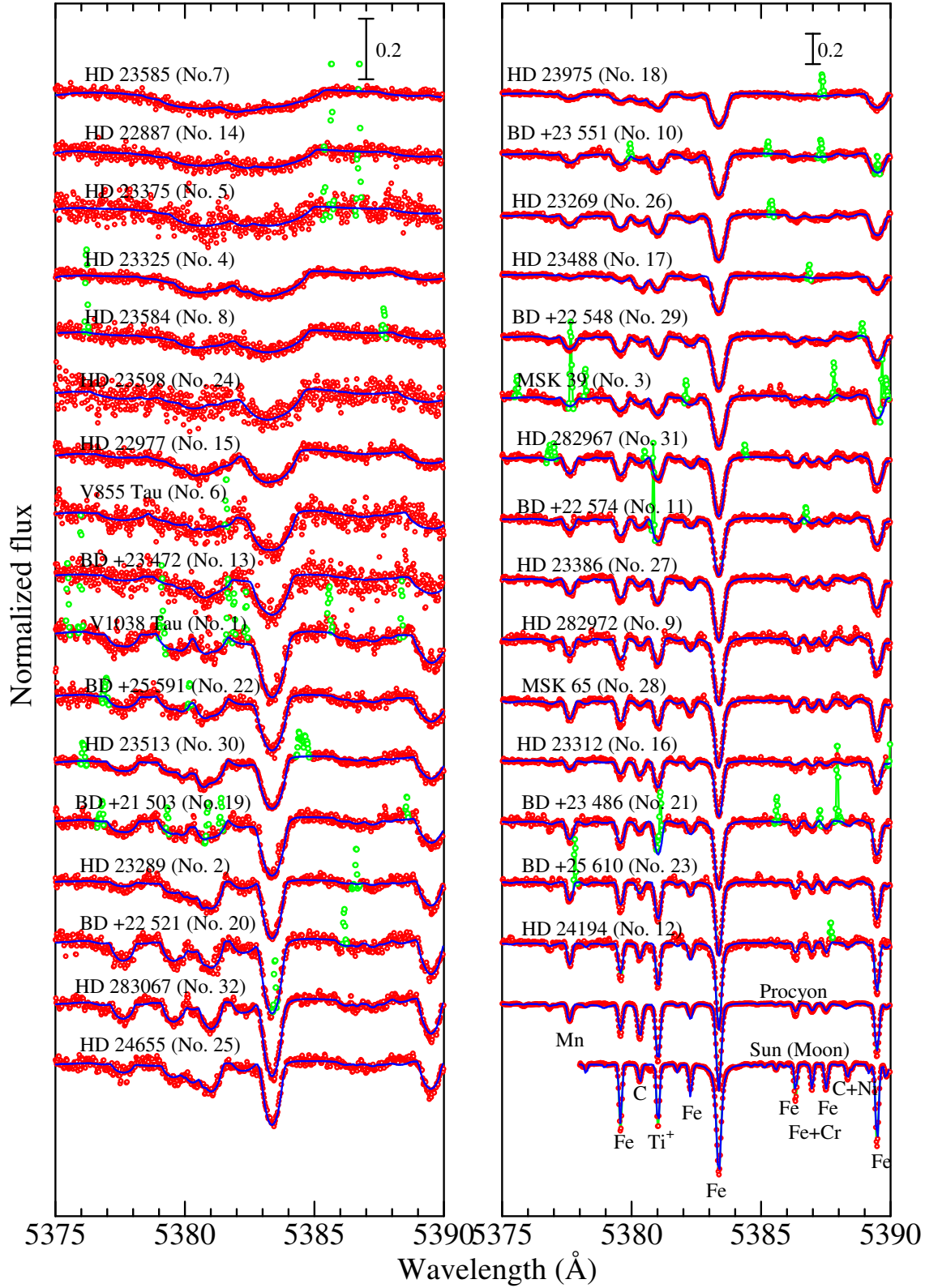}
  \end{center}
\caption{
Synthetic spectrum fitting for the 32 program stars (as well as Procyon
and the Sun/Moon) at the 5375--5390~$\rm\AA$ region, which was accomplished by 
adjusting the abundances of C, Ti, Mn, and Fe along with the macrobroadening 
parameter ($v_{\rm e} \sin i$). The best-fit theoretical spectra are shown by 
blue solid lines, while the observed data are plotted by red symbols 
(while those masked/disregarded in the fitting are highlighted in green).  
In each panel, the spectra are arranged in the descending order 
of $v_{\rm e}\sin i$ (cf. table~2), and an appropriate offset (0.2 and 
0.4 for the left and right panel, respectively) is applied to each 
spectrum relative to the adjacent one. Note that the vertical scale
in the left-had panel is twice as magnified as that of the right-hand 
panel. The wavelength scale is adjusted to the laboratory frame
by correcting the radial-velocity shift.
}
\end{figure}

\begin{figure}
  \begin{center}
    \FigureFile(150mm,200mm){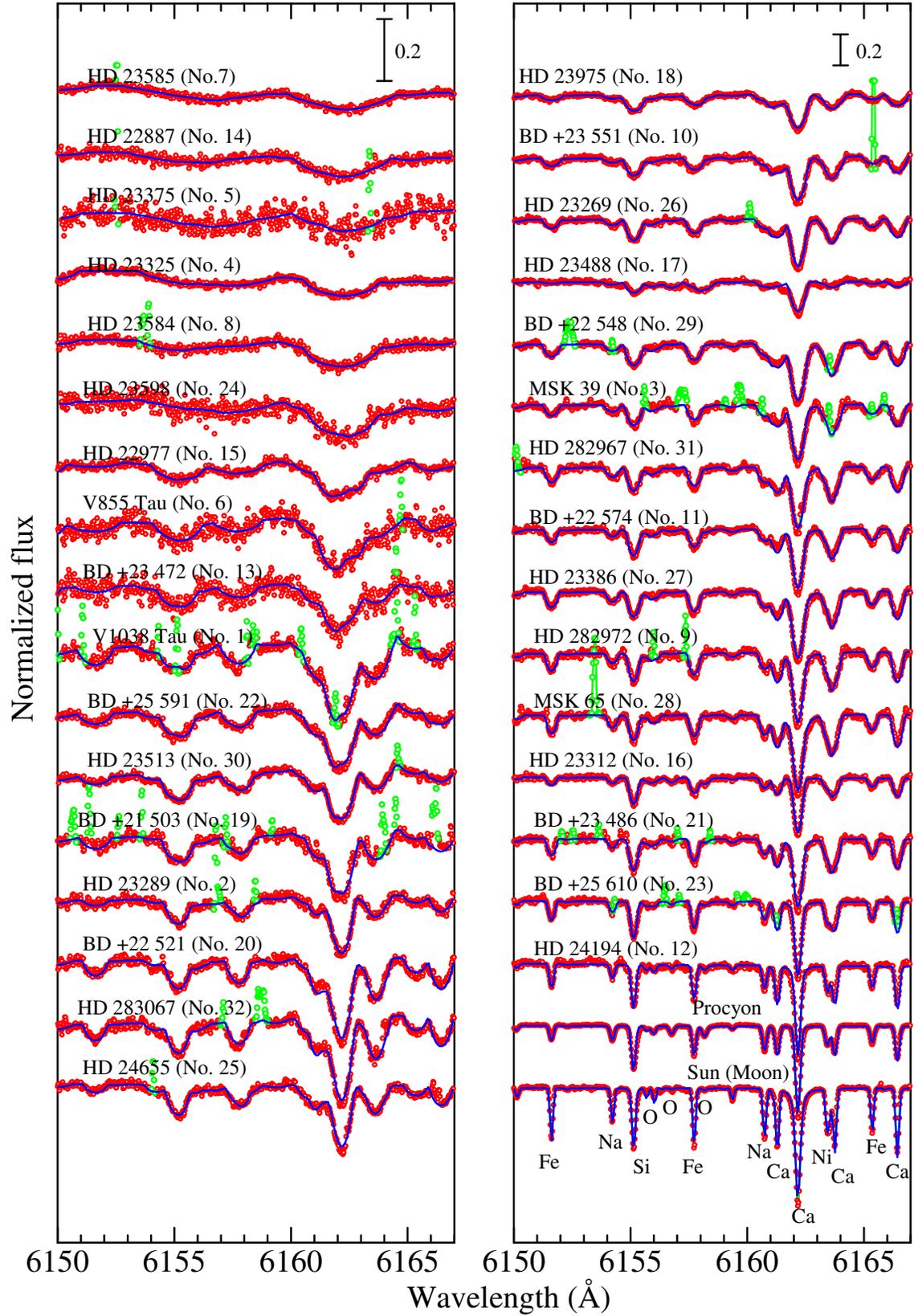}
  \end{center}
\caption{
Synthetic spectrum fitting for the 32 program stars (as well as Procyon and
the Sun/Moon) at the 6150--6167~$\rm\AA$ region, which was accomplished by 
adjusting the abundances of O, Na, Si, Ca, Fe, and Ni along with the 
macrobroadening parameter ($v_{\rm e} \sin i$). Otherwise, the same as 
in figure~3.
}
\end{figure}

\begin{figure}
  \begin{center}
    \FigureFile(100mm,150mm){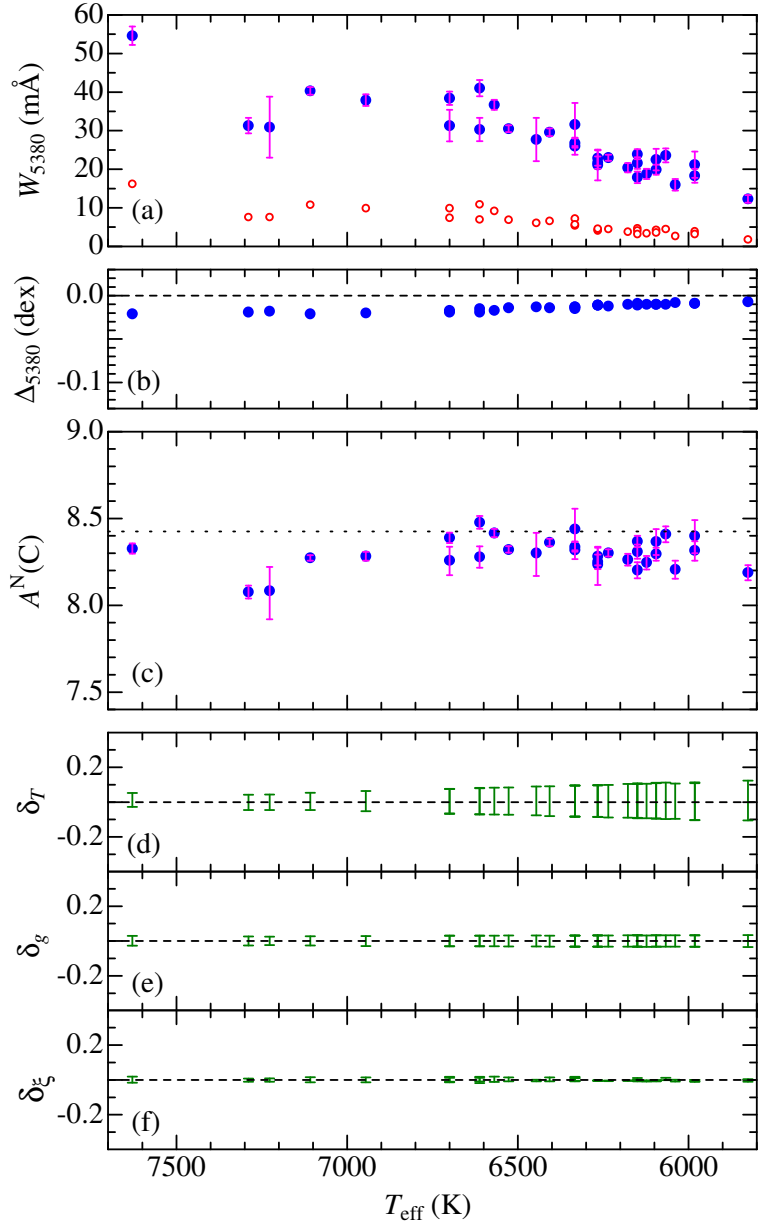}
  \end{center}
\caption{
C~{\sc i}~5380-related quantities plotted against $T_{\rm eff}$.
(a) $W_{5380}$ (equivalent width of the C~{\sc i} 5380.325 line
[filled symbols] and the C~{\sc i} 5380.227 line [open symbols] 
inversely computed with the  abundance solution derived from fitting), 
(b) $\Delta_{5380}$ (non-LTE correction for the C~{\sc i} 5380.325 line),
(c) $A^{\rm N}$(C) (non-LTE abundance derived from $W_{5380.325}$),  
(d) $\delta_{T\pm}$ (abundance changes in response to $T_{\rm eff}$ 
variations by $\pm$200~K), (f) $\delta_{g\pm}$ (abundance changes in
response to $\log g$ variations by $\pm$0.1~dex), and 
(g) $\delta_{\xi\pm}$ and (abundance change in response to  
$\xi$ variations by $\pm 0.5$~km~s$^{-1}$).
The signs of $\delta$'s concerning the variations of $T_{\rm eff}$, $\log g$, 
and $\xi$ are $\delta_{T+}<0$, $\delta_{T-}>0$, $\delta_{g+}>0$, 
$\delta_{g-}<0$, $\delta_{\xi +}<0$, and $\delta_{\xi -}>0$.
The error bars attached to each of the symbols in panels (a) and (c) 
denote the S/N-related uncertainties in equivalent widths ($\pm \delta W$) 
estimated by Cayrel's (1988) relation and the corresponding abundance 
errors ($\delta _{W+}$ and $\delta _{W-}$), respectively.
Note that the ordinate scale of panel (b) is as 5 times
expanded as that of panels (c)--(f). The dotted line in panel (c)
denotes the reference solar C abundance of 8.43 (cf. subsection~4.3).
}
\end{figure}

\begin{figure}
  \begin{center}
    \FigureFile(100mm,150mm){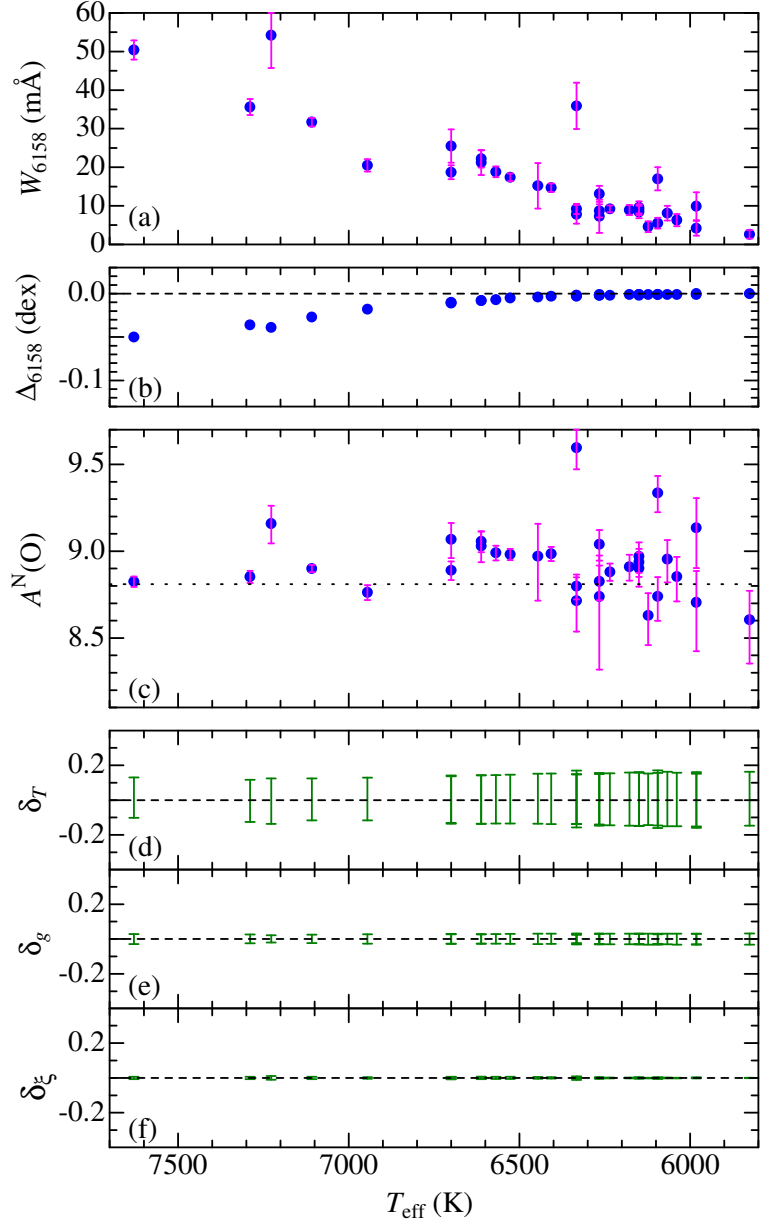}
  \end{center}
\caption{
O~{\sc i}~6158-related quantities plotted against $T_{\rm eff}$.
(a) $W_{6158}$ (equivalent width of the O~{\sc i} 6158 line (comprising 
3 components) inversely computed with the  abundance solution derived 
from fitting), (b) $\Delta_{6158}$ (non-LTE correction for O~{\sc i} 
6158), (c) $A^{\rm N}$(O) (non-LTE abundance derived from $W_{6158}$),  
(d) $\delta_{T\pm}$ (abundance changes in response to $T_{\rm eff}$ 
variations by $\pm$200~K), (f) $\delta_{g\pm}$ (abundance changes in
response to $\log g$ variations by $\pm$0.1~dex), and 
(g) $\delta_{\xi\pm}$ (abundance change in response to 
$\xi$ variations by $\pm 0.5$~km~s$^{-1}$).
The dotted line in panel (c) denotes the reference solar O abundance of 8.81
(cf. footnote~8).
Otherwise, the same as in figure~5.
}
\end{figure}

\begin{figure}
  \begin{center}
    \FigureFile(130mm,130mm){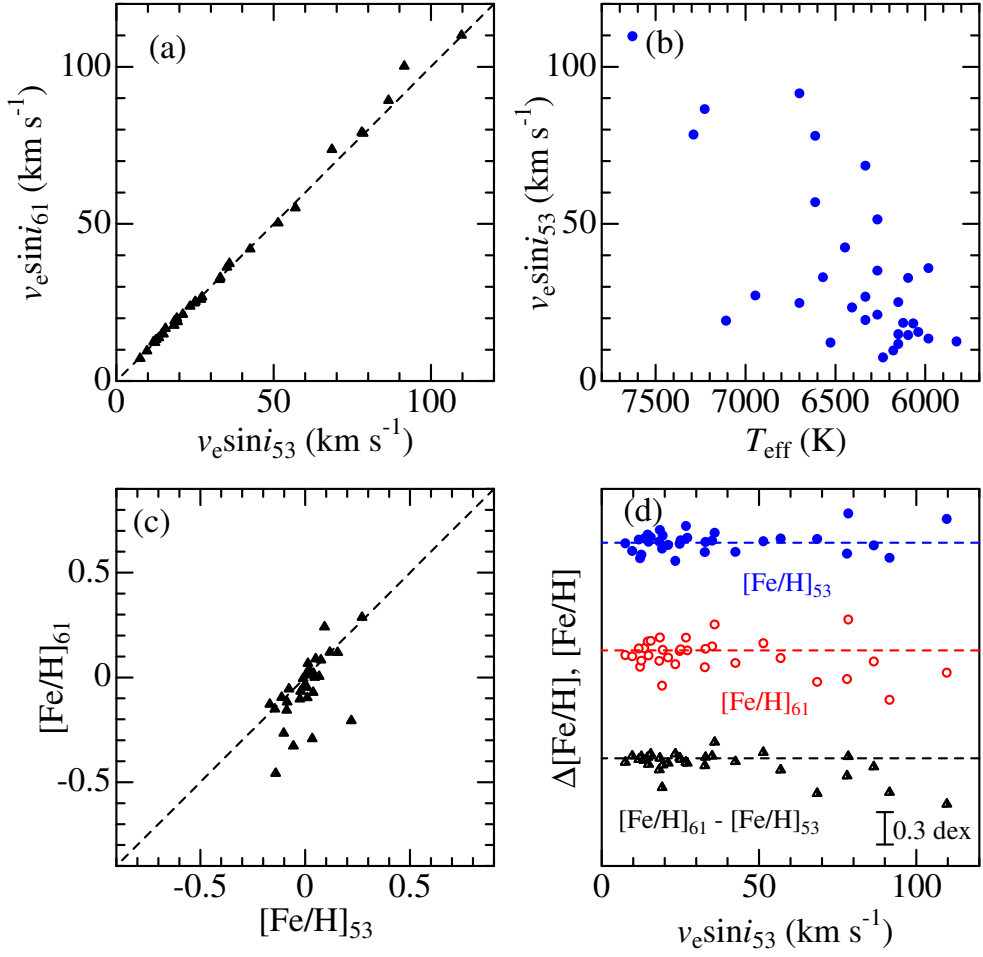}
  \end{center}
\caption{
(a) Comparison of the $v_{\rm e}\sin i$ values derived from 5375--5390~$\rm\AA$ 
fitting ($v_{\rm e}\sin i_{53}$; abscissa) with those from 6150--6167~$\rm\AA$ 
fitting ($v_{\rm e}\sin i_{61}$; ordinate). (b) $v_{\rm e}\sin i_{53}$ values
plotted against $T_{\rm eff}$. (c) Comparison of the [Fe/H] values derived 
from 5375--5390~$\rm\AA$ fitting ([Fe/H]$_{53}$; abscissa) with those from 
6150--6167~$\rm\AA$ fitting ([Fe/H]$_{61}$; ordinate). (d) [Fe/H]$_{53}$ (upper),
[Fe/H]$_{61}$ (middle), and their difference ([Fe/H]$_{61} - $[Fe/H]$_{53}$; lower)
plotted against $v_{\rm e}\sin i_{53}$. The horizontal dashed lines indicate 
the zero position for each case.
}
\end{figure}

\begin{figure}
  \begin{center}
    \FigureFile(100mm,150mm){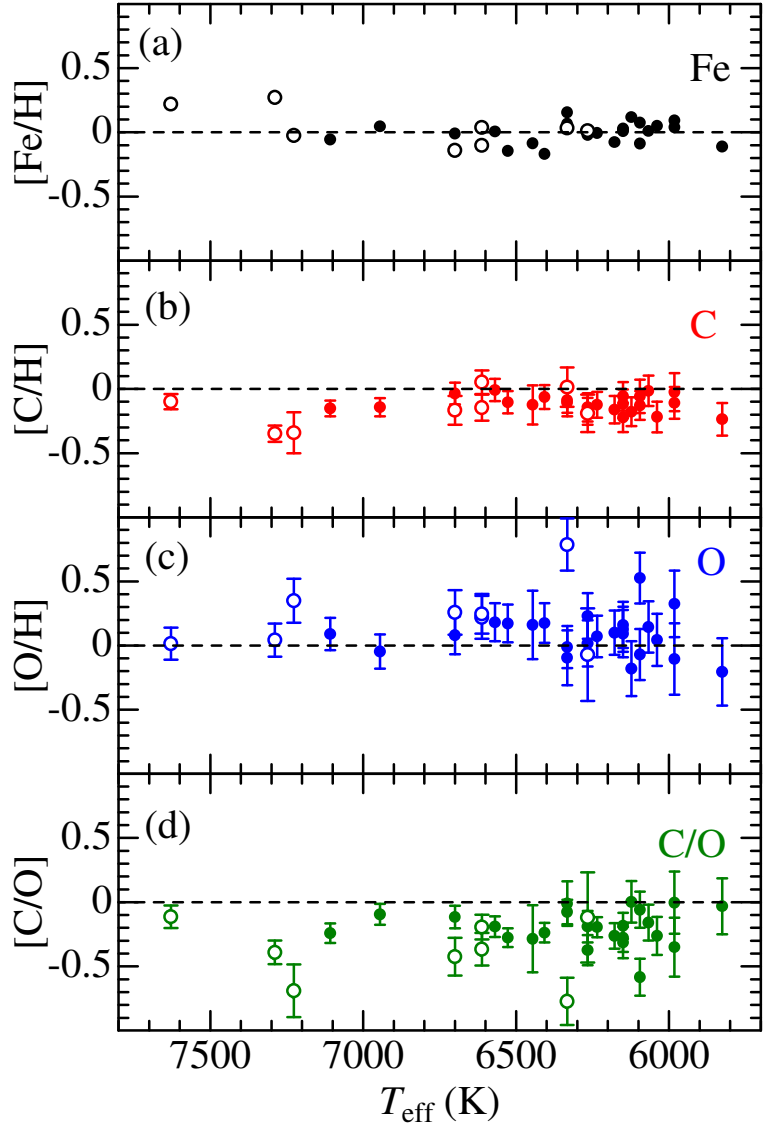}
  \end{center}
\caption{
Logarithmic relative abundances plotted against 
$T_{\rm eff}$. (a) [Fe/H], (b) [C/H], (c) [O/H], and (d) [C/O] 
($\equiv$ [C/H]~$-$~[O/H]), based on the data presented in table~2.
Filled circles denote stars with $v_{\rm e}\sin i < 50$~km~s$^{-1}$,
while open circles correspond to those with $v_{\rm e}\sin i > 50$~km~s$^{-1}$.
The error bar ($\pm \delta_{Tg\xi W}$) atached to each symbol in panels 
(b) and (c) is defined in terms of the quantities introduced in subsection~4.2 
(cf. the caption in figure~5)
as $\delta_{Tg\xi W} \equiv \sqrt{\delta_{T}^{2} + \delta_{g}^{2} + \delta_{\xi}^{2} + \delta _{W}^{2}}$,
where $\delta_{T} \equiv (|\delta_{T+}|+|\delta_{T-}|)/2$,
$\delta_{g} \equiv (|\delta_{g+}|+|\delta_{g-}|)/2$, 
$\delta_{\xi} \equiv (|\delta_{\xi+}|+|\delta_{\xi-}|)/2$), and
$\delta _{W} \equiv (|\delta _{W+}| + |\delta _{W-})|/2$. 
Meanwhile, the error bar ($\pm \Delta_{Tg\xi W}$) in panel (d) is differently
defined (because C/O ratio is concerned and parameter sensitivity tends
to be canceled) as:
$\Delta_{Tg\xi W} \equiv \sqrt{
(\delta^{\rm C/O}_{T})^{2} + (\delta^{\rm C/O}_{g})^{2} + (\delta^{\rm C/O}_{\xi})^{2}
+ (\delta^{\rm C}_{W})^{2} + (\delta^{\rm O}_{W})^{2},
}$,
where 
$\delta^{\rm C/O}_{T} \equiv (|\delta^{\rm C}_{T+}-\delta^{\rm O}_{T+}|+|\delta^{\rm C}_{T-}-\delta^{\rm O}_{T-}|)/2$,
$\delta^{\rm C/O}_{g} \equiv (|\delta^{\rm C}_{g+}-\delta^{\rm O}_{g+}|+|\delta^{\rm C}_{g-}-\delta^{\rm O}_{g-}|)/2$,
and
$\delta^{\rm C/O}_{\xi} \equiv (|\delta^{\rm C}_{\xi+}-\delta^{\rm O}_{\xi+}|+|\delta^{\rm C}_{\xi-}-\delta^{\rm O}_{\xi-}|)/2$.
}
\end{figure}

\begin{figure}
  \begin{center}
    \FigureFile(100mm,150mm){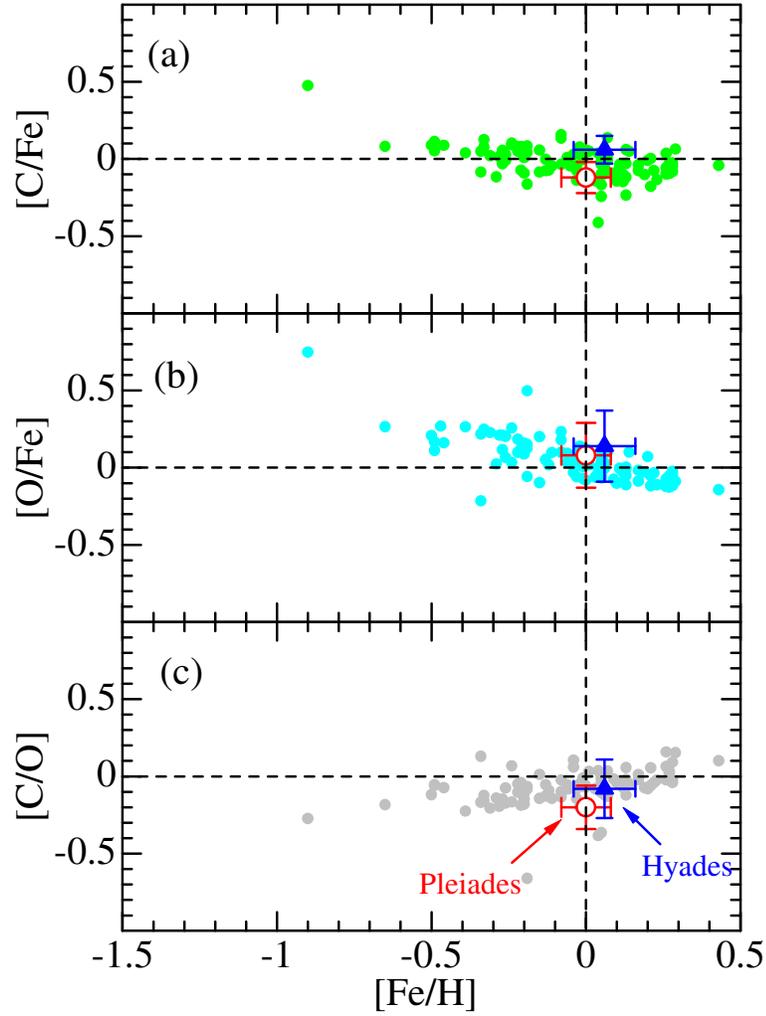}
  \end{center}
\caption{
Distributions of [C/Fe] (upper panel a), [O/Fe] (middle panel b), and 
[C/O] (lower panel c) plotted against [Fe/H] for Pleiades stars (mean; 
large open circle; from this study), Hyades stars (mean; large filed triangle;
from Paper~I) and field F--G stars (small filled circles; from Takeda \& Honda 2005).
In order to make the three data sets as consistent with each other as possible,
the data selections were done according to the following conditions: 
Pleiades data $\cdots$ stars for $v_{\rm e}\sin i < 50$~km~s$^{-1}$,
Hyades data $\cdots$ stars for $T_{\rm eff} > 5800$~K, and  field-star data $\cdots$ 
stars for $T_{\rm eff} > 5800$~K (only data based on C~{\sc i} 5380 and O~{\sc i} 6158 
lines were used). The error bars attached to Pleiades and Hyades data denote 
the standard deviations. 
}
\end{figure}

\begin{figure}
  \begin{center}
    \FigureFile(100mm,150mm){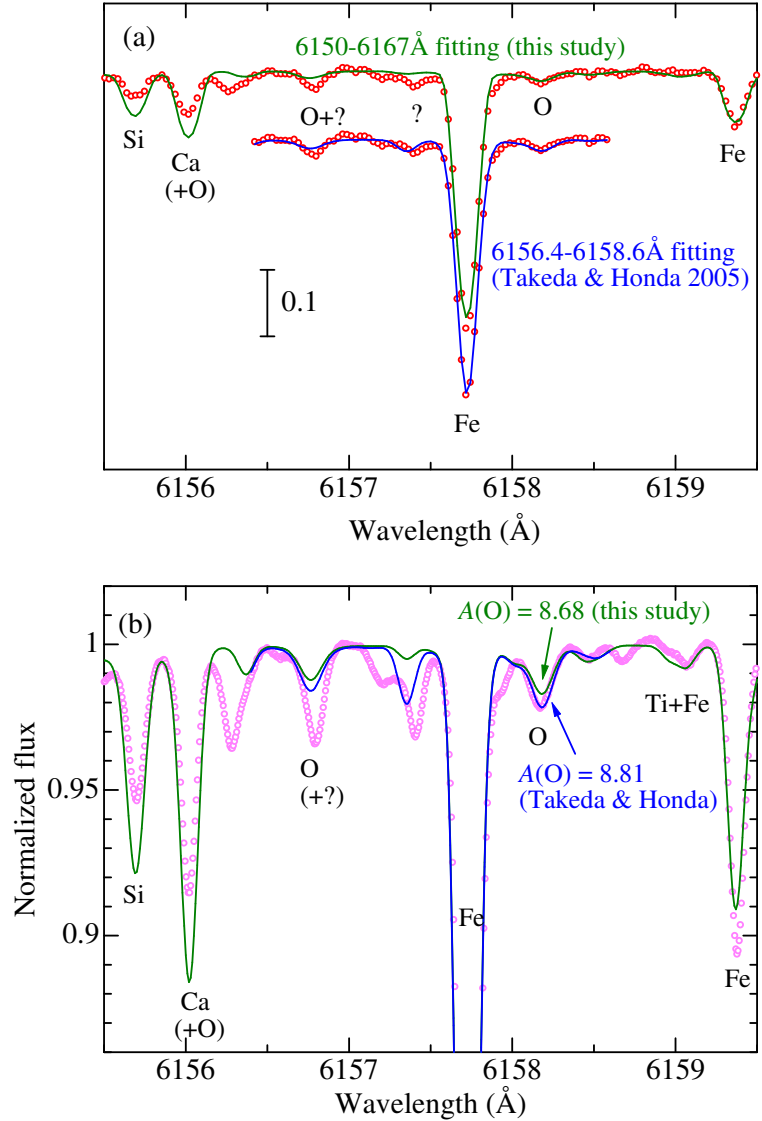}
  \end{center}
\caption{
(a) Comparison of two spectrum fittings in the region of O~{\sc i} 6156--8 lines
for solar O abundance determination done in this study (wide 6150--6167~$\rm\AA$ 
region) and in Takeda and Honda (2005) (narrow 6156.4--6158.6~$\rm\AA$ region 
with adjusted strengths of Ti lines to reproduce the 6157.4~$\rm\AA$ feature),
where open circles and lines represent the Moon spectrum (Takeda et al. 2005a) 
and the fitted theoretical spectra, respectively. 
(b) Simulated solar spectra (lines) corresponding to the abundance solutions
derived in this study and in Takeda and Honda (2005) (convolved with
a Gaussian function with an $e$-folding width of 2.5~km~s$^{-1}$) in comparison
of Kurucz et al.'s (1984) solar flux spectrum (its continuum position is 
slightly raised by 0.5\% in order to match the theoretical spectra).
}
\end{figure}

\begin{figure}
  \begin{center}
    \FigureFile(100mm,100mm){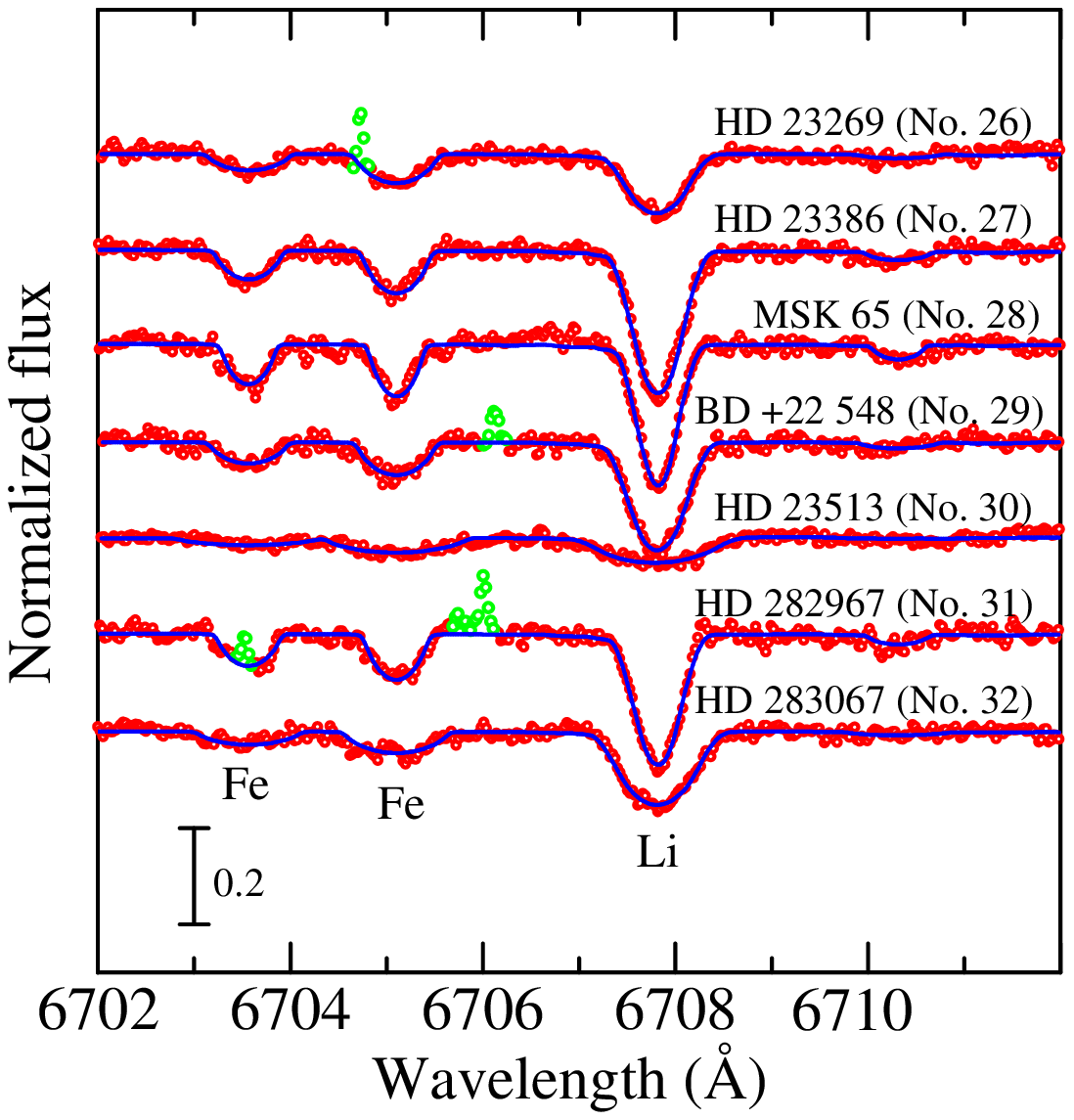}
  \end{center}
\caption{
Synthetic spectrum fitting at the 6702--6712~$\rm\AA$ region for 7 stars, for which 
spectra at the Li~{\sc i} 6708 line are available (i.e., GAOES data in the 2015--2016 season; 
cf. table~1). The best fit was accomplished by adjusting the abundances of Li and Fe 
along with the macrobroadening parameter ($v_{\rm e} \sin i$). The spectra are arranged
according to the object number (cf. table~1). An appropriate offset of 0.2 is applied 
to each spectrum relative to the adjacent one  Otherwise, the same as in figure~3.
}
\end{figure}


\begin{thebibliography}{}
\bibitem[]{}
  Allende Prieto, C., Barklem, P. S., Lambert, D. L., \& Cunha, K. 2004,
  A\&A, 420, 183
\bibitem[]{}
  Amarsi, A. M., Asplund, M., Collet, R., \& Leenaarts, J. 2016, MNRAS, 455, 3735
\bibitem[]{}
  Asplund, M., Grevesse, N., Sauval, A. J., \& Scott, P. 2009,
  ARA\&A, 47, 481
\bibitem[]{}
  Basri, G., Marcy, G. W., \& Graham, J. R. 1996, ApJ, 458, 600
\bibitem[]{}
  Bertran de Lis, S., Delgado Mena, E., Adibekyan, V. Kh, Santos, N. C.,
  \& Sousa, S. G. 2015, A\&A, 576, A89
\bibitem[]{}
  Boesgaard, A. M. 2005, in Cosmic Abundances as Records of Stellar Evolution
  and Nucleosynthesis in honor of David L. Lambert, eds. T. G. Barnes III
  and F. N. Bash,  ASP Conf. Ser., Vol. 336, p.39 (San Francisco, ASP) 
\bibitem[]{}
  Boesgaard, A. M., Armengaud, E., \& King, J. R. 2003, ApJ, 582, 410
\bibitem[]{}
  Bonatto, Ch, Bica, E., \& Girardi, L. 2004, A\&A, 415, 571
\bibitem[]{}
  Bond, J. C., O'Brien, D. P., \& Lauretta, D. S. 2010, ApJ, 715, 1050 
\bibitem[]{}
  Breger, M. 1986, ApJ, 309, 311
\bibitem[]{}
  Bressan, A., Marigo, P., Girardi, L., Salasnich, B., Dal Cero, C., Rubele, S., 
  Nanni, A. 2012, MNRAS, 427, 127
\bibitem[]{}
  Bressan, A., Marigo, P., Girardi, L., Nanni, A., \& Rubele, S. 2013, 
  EPJ Web of Conferences, 43, 3001 (DOI: http://dx.doi.org/10.1051/epjconf/20134303001)
\bibitem[]{}
  Carigi, L., Peimbert, M., Esteban, C., \& Garc\'{\i}a-Rojas, J. 2005, 
  ApJ, 623, 213
\bibitem[]{}
  Cayrel, R. 1988, in The Impact of Very High S/N Spectroscopy on Stellar Physics,
  Proc. IAU Symp. 132, eds. G. Cayrel de Strobel \& M. Spite (IAU), p. 345
\bibitem[]{}
  Delgado Mena, E., Israelian, G., Gonz\'{a}les Hern\'{a}ndez, J. I.,
  Bond, J. C., Santos, N. C., Udry, S., \& Mayor, M. 2010, ApJ, 725, 2349
\bibitem[]{}
  Edvardsson, B., Andersen, J., Gustafsson, B., Lambert, D. L., 
  Nissen, P. E., \& Tomkin, J. 1993, A\&A, 275, 101
\bibitem[]{}
  Esteban, C., Garc\'{\i}a-Rojas, J., Peimbert, M., Peimbert, A.,
  Ruiz, M. T., Rodr\'{\i}guez, M., \& Carigi, L. 2005, ApJ, 618, L95
\bibitem[]{}
  Friel, E. D., \& Boesgaard, A. M. 1990, ApJ, 351, 480
\bibitem[]{}
  Funayama, H., Itoh, Y., Oasa, Y., Toyota, E., Hashimoto, O., 
   \& Mukai, T. 2009, PASJ, 61, 931 
\bibitem[]{}
  Gebran, M., \& Monier, R. 2008, A\&A, 483, 567
\bibitem[]{}
  Georgy, C., Ekstr\"{o}m, S., Granada, A., Meynet, G., Mowlavi, N.,
  Eggenberger, P., \& Maeder, A. 2013, A\&A, 553, A24
\bibitem[]{}
  Gray, D. F. 2005, The Observation and Analysis of Stellar Photospheres, 3rd ed.
  (Cambridge: Cambridge University Press)
\bibitem[]{}
  Gustafsson, B., Karlsson, T., Olsson, E., Edvardsson, B., \& Ryde, N. 1999, A\&A, 342, 426
\bibitem[]{}
  Hertzsprung, E. 1947,  Annalen van de Sterrewacht te Leiden, 19, A1
\bibitem[]{}
  Kurucz, R. L. 1993, Kurucz CD-ROM, No. 13 (Harvard-Smithsonian Center
  for Astrophysics)
\bibitem[]{}
  Kurucz, R. L., \& Bell, B. 1995, Kurucz CD-ROM, No. 23 
  (Harvard-Smithsonian Center for Astrophysics)
\bibitem[]{}
  Kurucz, R. L., Furenlid, L., Brault, J., \& Testerman, L. 1984,
  Solar Flux Atlas from 296 to 1300~nm (Sunspot, New Mexico: National
  Solar Observatory)
\bibitem[]{}
  Leushin, V. V., \& Topil'skaya, G. P. 1987, Astrophysics, 25, 415
\bibitem[]{}
  Melis, C., Reid, M. J., Mioduszewski, A. J., Stauffer, J. R., \& Bower, G. C. 2014,
  Science, 340, 1029
\bibitem[]{}
  Nakajima, T., \& Sorahana, S.2016, ApJ, in press (arXiv: 1607.07528)
\bibitem[]{}
  Nieva, M.-F., \& Przybilla, N. 2012, A\&A, 539, A143
\bibitem[]{}
  Nissen, P. E. 1981, A\&A, 97, 145 
\bibitem[]{}
  Nissen, P. E., Chen, Y. Q., Carigi, L., Schuster, W. J., \& Zhao, G. 2014, A\&A, 568, A25
\bibitem[]{}
  Pereira, T. M. D., Asplund, M., \& Kiselman, D. 2009, A\&A, 508, 1403
\bibitem[]{}
  Perryman, M. A. C., et al. 1998, A\&A, 331, 81
\bibitem[]{}
  Petigura, E. A., \& Marcy, G. W. 2011, ApJ, 735, 41
\bibitem[]{}
  Ryabchikova, T., Piskunov, N., Kurucz, R. L., Stempels, H. C., Heiter, U., Pakhomov, Yu, 
  \& Barklem, P. S. 2015, Physica Scripta, Vol. 90, Issue 5 
\bibitem[]{}
  Steffen, M. 1985, A\&AS, 59, 403
\bibitem[]{}
  Takeda, Y. 1995, PASJ, 47, 287
\bibitem[]{}
  Takeda, Y. 2007, PASJ, 59, 335
\bibitem[]{}
  Takeda, Y., et al. 2005a, PASJ, 57, 13
\bibitem[]{}
  Takeda, Y., Han, I., Kang, D.-I., Lee, B.-C., \& Kim, K.-M. 2008,
  JKAS, 41, 83
\bibitem[]{}
  Takeda, Y., \& Honda, S. 2005, PASJ, 57, 65
\bibitem[]{}
  Takeda, Y., \& Honda, S. 2015, PASJ, 67, 25
\bibitem[]{}
  Takeda, Y., Honda, S., Ohnishi, T., Ohkubo, M, Hirata, R., \& Sadakane, K. 2013,
  PASJ, 65, 53 (Paper~I)
\bibitem[]{}
  Takeda, Y., \& Kawanomoto, S. 2005, PASJ, 57, 45 
\bibitem[]{}
  Takeda, Y., Ohkubo, M., Sato, B., Kambe, E., \& Sadakane, K. 2005b, PASJ, 57, 27
  [Erratum: PASJ, 57, 415]
\end{thebibliography}
\end{document}